\documentclass[
  journal=pasa,
  manuscript=research-paper, 
  year=2020,
  volume=37,
]{cup-journal}

\usepackage{microtype,siunitx,booktabs}
\usepackage{multirow}
\usepackage{pdflscape}
\usepackage{longtable}
\usepackage{caption}
\usepackage{subcaption}
\DeclareCaptionFormat{cont}{#1 (cont.)#2#3\par}

\usepackage{hyperref}
\hypersetup{colorlinks,citecolor=blue,linkcolor=blue,urlcolor=blue}

\usepackage{array}
\newcolumntype{P}[1]{>{\raggedright\arraybackslash}p{#1}}
\newcolumntype{A}[1]{>{\raggedleft\arraybackslash}p{#1}}

\def\arcsec{\hbox{$^{\prime\prime}$}}

\newcommand\farcs{\hbox{$.\!\!^{\prime\prime}$}}

\newcommand{\gaia}{\textit{Gaia}}

\newcommand{\asec}{$^{\prime\prime}$}

\newcommand*\oline[1]{%
  \vbox{%
    \hrule height 0.5pt
    \kern0.25ex
    \hbox{%
      \kern-0.1em
      \ifmmode#1\else\ensuremath{#1}\fi
      \kern-0.1em
    }
  }
}  

\usepackage{xcolor}

\usepackage{amsmath}	
\usepackage{amssymb}	

\title{Detection of radio emission from stars via proper-motion searches}

\author{Laura N. Driessen}
\affiliation{Sydney Institute for Astronomy, School of Physics, University of Sydney, NSW 2006, Australia\\
}
\alsoaffiliation{CSIRO, Space and Astronomy, PO Box 1130, Bentley, WA 6102, Australia\\}
\email[L. N. Driessen]{Laura@Driessen.net.au}

\author{George Heald}
\affiliation{CSIRO, Space and Astronomy, PO Box 1130, Bentley, WA 6102, Australia\\}

\author{Stefan W. Duchesne}
\affiliation{CSIRO, Space and Astronomy, PO Box 1130, Bentley, WA 6102, Australia\\}

\author{Tara Murphy}
\affiliation{Sydney Institute for Astronomy, School of Physics, University of Sydney, NSW 2006, Australia\\
}
\alsoaffiliation{ARC Centre of Excellence for Gravitational Wave Discovery (OzGrav), Hawthorn, Victoria, Australia\\}

\author{Emil Lenc}
\affiliation{CSIRO, Space and Astronomy, PO Box 76, Epping, NSW 1710, Australia\\}

\author{James K. Leung}
\affiliation{Sydney Institute for Astronomy, School of Physics, University of Sydney, NSW 2006, Australia\\
}
\alsoaffiliation{CSIRO, Space and Astronomy, PO Box 76, Epping, NSW 1710, Australia\\}
\alsoaffiliation{ARC Centre of Excellence for Gravitational Wave Discovery (OzGrav), Hawthorn, Victoria, Australia\\}

\author{Vanessa A. Moss}
\affiliation{CSIRO, Space and Astronomy, PO Box 76, Epping, NSW 1710, Australia\\}
\alsoaffiliation{Sydney Institute for Astronomy, School of Physics, University of Sydney, NSW 2006, Australia\\
}

\doi{10.1017/pasa.2020.32}

\received {dd Mmm YYYY}
\revised  {dd Mmm YYYY}
\accepted {dd Mmm YYYY}
\published{22 September 2020}

\keywords{radio astrometry;
radio continuum emission;
Galactic radio sources;
proper motions;
stellar flares;} 

\begin{document}

\begin{abstract}
We present a method for identifying radio stellar sources using their proper-motion. We demonstrate this method using the FIRST, VLASS, RACS-low and RACS-mid radio surveys, and astrometric information from \gaia\, Data Release 3. We find eight stellar radio sources using this method, two of which have not previously been identified in the literature as radio stars. We determine that this method probes distances of $\sim90$pc when we use FIRST and RACS-mid, and $\sim250$pc when we use FIRST and VLASS.
We investigate the time baselines required by current and future radio sky surveys to detect the eight sources we found, with the SKA (6.7 GHz) requiring $<3$ years between observations to find all eight sources. 
We also identify {nine previously known and $43$} candidate variable radio stellar sources that are detected in FIRST (1.4 GHz) but are not detected in RACS-mid (1.37 GHz). This shows that many stellar radio sources are variable, and that surveys with multiple epochs can detect a more complete sample of stellar radio sources.
\end{abstract}

\section{Introduction}
\label{sec: introduction}

A recent large increase in the sample of known radio stars \citep[see e.g.][for a catalogue of stellar radio sources]{1995A&AS..109..177W} has been enabled by the advent of wide-field of view, high resolution interferometers such as the Karl G. Jansky Very Large Array \citep[VLA;][]{perley2011}, the Low Frequency Array \citep[LOFAR;][]{2013A&A...556A...2V}, the Australian Square Kilometre Array Pathfinder
\citep[ASKAP\footnote{\href{https://www.atnf.csiro.au/projects/askap/index.html}{https://www.atnf.csiro.au/projects/askap/index.html}};][]{2021PASA...38....9H} and the (more) Karoo Array Telescope \citep[MeerKAT;][]{2018ApJ...856..180C}. New radio stars have been identified using circular polarisation searches \citep[e.g.][]{2021MNRAS.502.5438P,2021NatAs...5.1233C,2021A&A...654A..21T},
and variability searches \citep[e.g.][]{2020MNRAS.491..560D,2022MNRAS.510.1083D,2022MNRAS.513.3482A}. Finding radio stars is important for probing the physics behind stellar radio emission, for searching for radio signatures of exoplanets {\citep{2009ApJ...701.1922B,2020AJ....160...97C}}, and for tying optical and radio reference frames together. As we use current instruments and look forward to the SKA, we need to consider new methods for searching for and confirming the detection of radio emission from stellar sources.

The key challenge of identifying stellar radio emission is chance coincidence with background radio galaxies. Direct position matches between the optical and radio result in high chance coincidence probability \citep{2019RNAAS...3...37C} unless the samples are first restricted using the physical properties of the radio and/or optical sources. For example, both ASKAP and LOFAR have been used to perform circular polarisation searches for stellar sources \citep[e.g.][]{2021MNRAS.502.5438P,2021NatAs...5.1233C,2021A&A...654A..21T}. In the radio sky, only pulsars and stellar radio sources are known to have {high} circular polarisation fractions, and the sky density of {highly} circularly polarised sources is low. This reduces the chance coincidence probability and provides a physical reason to believe that a match between a {highly} circularly polarised radio source and an optical star is true. This has been demonstrated with great success by \citet{2021MNRAS.502.5438P} using ASKAP where they identified 10 known and 23 previously unknown radio stars, and by \citet{2021NatAs...5.1233C} and \citet{2021A&A...654A..21T} using LOFAR where they detected 1 known and 18 previously unknown active stars, and 14 known RS Canum Venaticorum (RS CVn) respectively. {Highly} circularly polarised stellar emission is typically coherent and non-thermal, caused by either plasma emission or electron-cyclotron maser emission \citep[see e.g.][for a review of stellar radio emission mechanisms]{1985ARA&A..23..169D}. This means that circular polarisation searches are biased towards coherent emission processes.

Stellar systems are known to flare in the radio on typical time scales of minutes to hours \citep[see][for a summary]{2008arXiv0801.2573O}, which means stellar systems can be found in radio variability searches. 
{All searches for radio stars are biased towards stars that flare more often as we are more likely to detect such stars. This is particularly the case for the MeerKAT results due to the serendipitous nature of the discoveries in variability searches \citep{2020MNRAS.491..560D,2022MNRAS.510.1083D,2022MNRAS.513.3482A}.}
{A search method that does not rely on high circular polarisation fraction and can be used for both quiescent and flaring stars would reduce the current biases in searches for radio emission from stellar sources.}

Very-Long Baseline Interferometry (VLBI) of stellar radio sources has been used in the past to perform high-precision radio astrometry. This has been done for various reasons including: astrometric monitoring to search for signatures of orbiting exoplanets \citep[e.g.][]{1994Ap&SS.212..251L,1995ASPC...74..219J}, to link optical and radio reference frames \citep[e.g.][]{1988AJ.....96.1746L,1999A&A...344.1014L}, and to distinguish between background galaxies and foreground radio stars using the stars' proper-motion \citep{1992A&A...258..112L}. However, searching for proper-motion radio stars has not previously been expanded to large-scale radio surveys. We now have long time baselines between the VLA Faint Images of the Radio Sky at Twenty-centimetres \citep[FIRST;][performed between 1993 and 2011]{1994ASPC...61..165B,1995ApJ...450..559B}, the VLA Sky Survey \citep[VLASS;][performed between 2017 and 2019]{2021ApJS..255...30G}, and the Rapid ASKAP Continuum Survey (RACS\footnote{ASKAP data, including the RACS-low and RACS-mid data (DOI: \href{https://doi.org/10.25919/1khs-c716}{https://doi.org/10.25919/1khs-c716}), are publicly available and can be accessed via the CSIRO ASKAP Science Data Archive (CASDA): \href{https://research.csiro.au/casda/}{https://research.csiro.au/casda/}.}) at 887.5 MHz \citep[RACS-low;][performed between 2019 and 2020]{2020PASA...37...48M,2021PASA...38...58H}, and at 1367.5 MHz (RACS-mid; Duchesne et al. submitted, performed between 2020 and 2022).
We also have high-precision proper-motion measurements from \gaia\,Data Release 3 \citep[DR3;][]{2016A&A...595A...1G,2022arXiv220605989B,2022gdr3.reptE..19V}. We can now combine these surveys to search for radio stars using their proper-motion.

We have performed a search for stellar radio sources using the FIRST, VLASS, RACS-low, and RACS-mid radio surveys and the proper-motion properties of optical sources from \gaia\,DR3. In Section\,\ref{sec: method} we present the search method. In Section\,\ref{sec: demo results} we present the results of using the method with FIRST, VLASS and RACS. In Section\,\ref{sec: demo unmatched} we present a search for candidate variable radio stellar sources. In Sections\,\ref{sec: discussion} and \ref{sec: conclusions} we discuss the results and conclude.


\section{Method}
\label{sec: method}

We used the proper-motion information from \gaia\,DR3 and the position of radio sources in FIRST, VLASS, RACS-low and RACS-mid to search for radio stars.

The FIRST survey was performed using the VLA in B-configuration at 1400 MHz between 1993 and 2011. It covers over 10\,000 square degrees and has a minimum declination of $\sim-10^{\circ}$.
It has an astrometric accuracy of 1\arcsec\, and a typical root-mean-square (RMS) noise of $\sim0.2$ mJy \citep[][]{1994ASPC...61..165B}. VLASS was performed using the VLA in B- and BnA-configuration at 2000--4000 MHz between 2017 and 2019. It covers the entire sky above a declination of $\sim-40^{\circ}$.
It has an astrometric accuracy of 0\farcs5\, above $\sim-20^{\circ}$.
Epoch 1.1 of VLASS has a typical RMS noise of 128 $\mu$Jy beam$^{-1}$ and epoch 1.2 has a typical RMS noise of 145 $\mu$Jy beam$^{-1}$.
RACS was performed using the 36-antenna ASKAP telescope with as many antennas as available at any given point. RACS-low was observed at 887.5 MHz between 2019 and 2020 and RACS-mid was observed at 1367.5 MHz between 2020 and 2022. Both RACS surveys have an astrometric accuracy of 2\arcsec. RACS-low and RACS-mid have median RMS noise across all tiles of $\sim0.27$ mJy beam$^{-1}$ and $\sim0.20$ mJy beam$^{-1}$ respectively. 
We create a simple, alternate catalogue for RACS-mid by concatenating source-lists from the individual RACS-mid images. Relevant metadata, such as observation start time, are added to each source entry during this process. There are a total of 4\,944\,458 sources in the concatenated catalogue. As each RACS-mid observation overlaps with adjacent observations to provide uniform sensitivity over the full survey, there are $\gtrsim 700\,000$ sources recorded more than once.
A summary of the radio survey details is shown in Table\,\ref{tab: survey info}. \gaia\, is a European Space Agency (ESA) space observatory that has been designed to measure precise positions, distances and proper-motions of optical sources. The third data release was made available on 2022 June 13 and contains astrometric information (and more) for $\sim1.46\times10^{9}$ sources.

\begin{table*}
\caption{Survey information for FIRST, RACS-mid, RACS-low and VLASS. The total number of sources includes all of the sources without any cuts applied; except for VLASS, which has been restricted to sources where the MainSample flag $=1$. {The integration time is the typical time per pointing. As VLASS is observed ``on the fly'', it does not have a typical integration time. Instead, VLASS was observed at a rate of approximately $23.83\,\mathrm{arcmin\,hour^{-1}}$ \citep{2020PASP..132c5001L}.}}
    \centering
    \begin{tabular}{lA{1.8cm}rrA{1.8cm}rA{1.5cm}A{1.5cm}A{1.5cm}}
     Survey     & Frequency range (MHz)    & Earliest epoch & Latest epoch & Typical RMS (mJy beam$^{-1}$) & Number of sources & Position accuracy (\arcsec) & Declination range & Integration time (s)\\
     \hline
     FIRST      & 1354.5 -  1445.5         & J1993.207 & J2011.312 & 0.2   & 946\,432    & 1.0           & $>-11.5^{\circ}$ & 180\\
     RACS-low   & 743.5 - 1031.5      & J2019.302 & J2020.472 & 0.27  & 2\,665\,933   & $\sim2.0$     & $<30^{\circ}$ & 900\\
     RACS-mid   & 1295.5 - 1439.5        & J2020.969 & J2022.173 & 0.2 & 4\,944\,458   & $\sim2.0$     & $<45^{\circ}$ & 900\\
     VLASS      & 2000 - 4000     & J2017.685 & J2019.539 & 0.13  & 1\,880\,195   & 0.5           & $>-20^{\circ}$ & \\
    \end{tabular}
    \label{tab: survey info}
 \end{table*}

To perform the proper-motion matching, we determined the positions at epoch A and epoch B of an optical source that has a proper-motion.
If the epoch A position of the optical source matched the position of a radio source from Survey A observed on epoch A, and the epoch B position of the optical source matched the position of a radio source from Survey B observed on epoch B, then the optical source is the radio source.
There are some caveats to this simple matching.

\begin{figure}
\caption{Diagram showing the definitions of the separations between the radio and \gaia\,positions.
We will be discussing positions and separations in different epochs and comparing radio and optical positions.
In this diagram we define: $D_\mathrm{{G_{A,B}}}$ as the separation between the \gaia\,position proper-motion corrected to epoch A ($G_{A}$) and the \gaia\,position proper-motion corrected to epoch B ($G_{B}$); $D_{\mathrm{R_{A}R_{B}}}$ as the separation between the survey A radio source position ($R_{A}$) and the survey B radio source position ($R_{B}$); $D_\mathrm{{G_{A}R_{A}}}$ as the separation between the \gaia\,position proper-motion corrected to epoch A and the survey A radio position; and $D_\mathrm{{G_{B}R_{B}}}$ as the separation between the \gaia\,position proper-motion corrected to epoch B and the survey B radio position.
}
\includegraphics[width=\columnwidth]{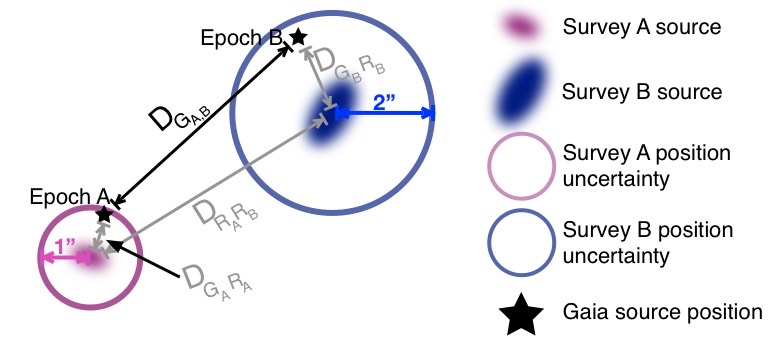}
\label{fig: radio D labels}
\end{figure}

We assumed that we had two radio surveys, survey A and survey B, where survey A has position accuracy $a$\arcsec\,and is earlier (epoch A)  than survey B (epoch B) with position accuracy $b$\arcsec, illustrated in Figure\,\ref{fig: radio D labels}. We defined the requirements for a proper-motion match between source A from survey A and source B from survey B to be:
\begin{enumerate}
    \item \label{req: F int F peak} both source A and source B have $F_{\mathrm{int}}/F_{\mathrm{peak}}\leq 1.5$ where $F_{\mathrm{int}}$ and $F_{\mathrm{peak}}$ are the integrated and peak flux densities respectively
    \item \label{req: A sep from all B} source A is separated by $D_\mathrm{{R_{A}R_{B}}}>a\arcsec+b\arcsec$ from any and all survey B sources
    \item \label{req: B sep from all A} source B is separated by $D_\mathrm{{R_{A}R_{B}}}>a\arcsec+b\arcsec$ from any and all survey A sources
    \item \label{req: A match to Gaia} the source A position and the \gaia\,position proper-motion corrected to epoch A are separated by $D_\mathrm{{G_{A}R_{A}}}<a\arcsec$
    \item \label{req: B match to Gaia} the source B position and the \gaia\,position proper-motion corrected to epoch B are separated by $D_\mathrm{{G_{B}R_{B}}}<b\arcsec$
    \item \label{req: A sep from Gaia B} the survey A position and the \gaia\,position proper-motion corrected to epoch B are separated by $D_\mathrm{{G_{B}R_{A}}}>a\arcsec$
    \item \label{req: B sep from Gaia A} the survey B position and the \gaia\,position proper-motion corrected to epoch A are separated by $D_{\mathrm{G_{A}R_{B}}}>b\arcsec$
\end{enumerate}

\begin{figure}
\caption{Diagram illustrating which radio sources are kept and which radio sources are removed to satisfy requirements\,\ref{req: A sep from all B} and \ref{req: B sep from all A}. All of the sources from both surveys within the black box are discarded while the sources outside of the box are kept. 
}
\includegraphics[width=\columnwidth]{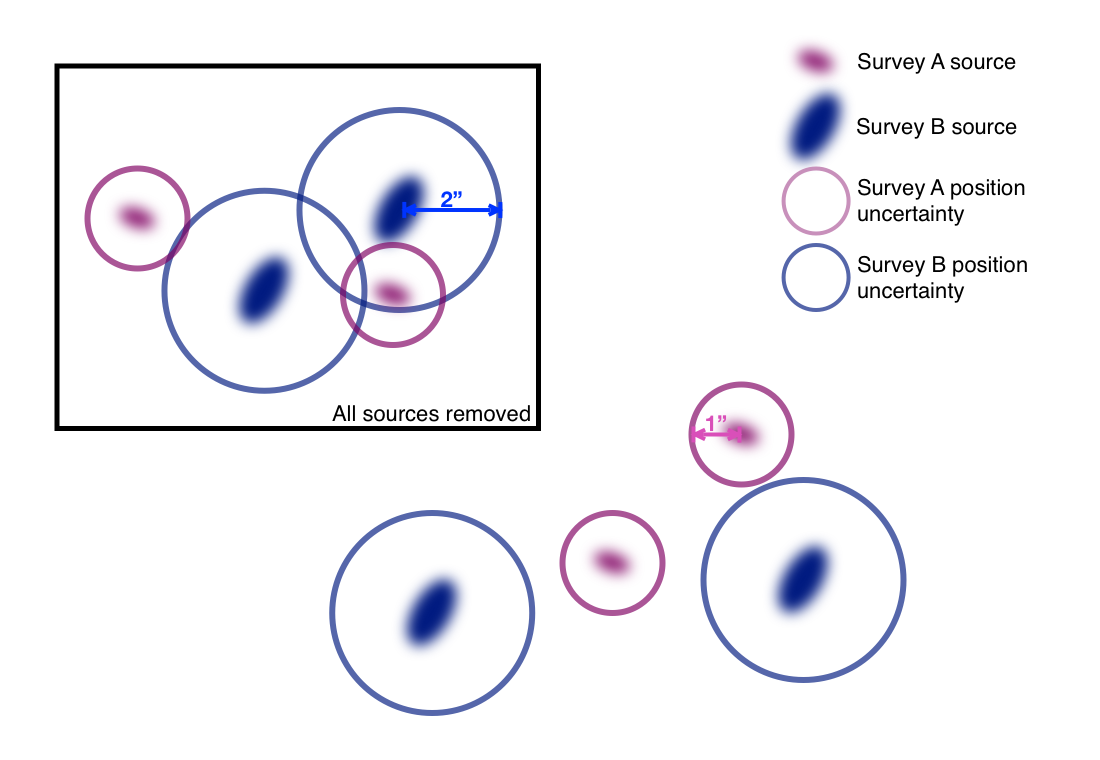}
\label{fig: radio remove}
\end{figure}

Requirement \ref{req: F int F peak} is to remove resolved sources from the set of survey A and B sources.
Requirements \ref{req: A sep from all B}, \ref{req: B sep from all A}, \ref{req: A sep from Gaia B} and \ref{req: B sep from Gaia A} are to ensure that the radio source is associated with the optical high-proper-motion object, instead of a background source that does not have a proper-motion. A diagram illustrating which radio sources would be removed to satisfy requirements \ref{req: A sep from all B} and \ref{req: B sep from all A} is shown in Figure\,\ref{fig: radio remove}. 
Requirements \ref{req: A match to Gaia} and \ref{req: B match to Gaia} ensure that the radio sources match the proper-motion source.

\begin{figure}
\caption{Diagram demonstrating a \gaia\,source that would be considered a radio proper-motion source. The black star indicates the position of the same \gaia\,source in the two different radio survey epochs.
}
\includegraphics[width=\columnwidth]{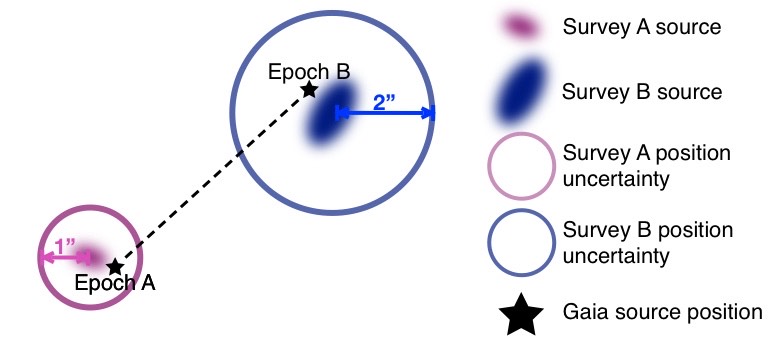}
\label{fig: Gaia radio pm source diagram}
\end{figure}

Practically, satisfying some of these requirements ensured that others are satisfied. First, we discarded the survey A and B sources where $F_{\mathrm{int}}/F_{\mathrm{peak}}\geq 1.5$ to satisfy requirement \ref{req: F int F peak}. If we found the separation between the remaining sources in survey A and survey B and removed sources in survey A that are $D_{\mathrm{R_{A}R_{B}}}<a\arcsec+b\arcsec$ from a survey B source and vice versa, we satisfied requirements \ref{req: A sep from all B} and \ref{req: B sep from all A}. If we proper-motion corrected the \gaia\,positions and matched them to survey A and discarded the \gaia\,sources where the separation is $>a\arcsec$, we satisfied \ref{req: A match to Gaia}. If we then proper-motion corrected the remaining \gaia\,sources with survey B and discarded \gaia\,sources where the separation is $>b\arcsec$, we satisfied requirement \ref{req: B match to Gaia}. The remaining \gaia\,sources necessarily satisfied requirements \ref{req: A sep from Gaia B} to \ref{req: B sep from Gaia A} as the survey A and B sources they were matched to are separated by $>a\arcsec+b\arcsec$. This means that our steps were as follows:
\begin{enumerate}
    \item \label{step: cut Gaia pm} {Discard \gaia\,sources that have no measured proper-motion magnitude}
    \item \label{step: resolved cut} Discard sources in survey A and survey B where $F_{\mathrm{int}}/F_{\mathrm{peak}}> 1.5$
    \item \label{step: discard A B} Keep any A sources that are separated by $>a\arcsec+b\arcsec$ from all B sources and keep any B sources that are separated by $>a\arcsec+b\arcsec$ from all A sources
    \item \label{step: xmatch A Gaia} Proper-motion correct the \gaia\,source positions to the survey A epoch and cross-match the source positions. The sources are considered a match if the separation is $<a\arcsec$. Discard those \gaia\,and survey A sources that do not match.
    \item \label{step: xmatch B Gaia} Proper-motion correct the remaining \gaia\,source positions to the survey B epoch and cross-match the source positions. The sources are considered a match if the separation is $<b\arcsec$. Discard those \gaia\,(and the corresponding survey A sources) and survey B sources that do not match.
\end{enumerate}
The remaining survey A, survey B and \gaia\,sources were considered proper-motion matches, a diagram demonstrating a \gaia\,source that meets our requirements is shown in Figure\,\ref{fig: Gaia radio pm source diagram}. We then manually examined the radio sources in the images to confirm the results and to confirm that the radio sources are unresolved.

We used FIRST as the oldest survey/survey A in all cases, and matched to each of VLASS, RACS-low and RACS-mid.
We used FIRST as it covers a large area of sky and has good position accuracy ($\sim1\arcsec$ for point sources). RACS-low and RACS-mid are the first southern hemisphere all-sky surveys with position accuracy ($\sim2\arcsec$) comparable to FIRST.
Both RACS surveys and VLASS were performed $\gtrsim25$ years after the first FIRST observations, providing an excellent time baseline, which meant that we could search for optical sources with lower proper-motions. The position accuracy, RMS noise level, declination range, number of sources and observing epoch ranges for FIRST, VLASS, RACS-low and RACS-mid are shown in Table\,\ref{tab: survey info}. We note that these surveys are not performed at the same frequencies. We therefore assumed that the stellar sources have close-to flat radio spectra. As per the VLASS quick look catalogue paper suggestions \citep{2021ApJS..255...30G}, we used the sources where the MainSample (or e.1) flag is 1. This meant that only sources with declination $>-20^{\circ}$ were included.

\section{Results}
\label{sec: demo results}

We performed the matches using the radio position accuracies shown in Table\,\ref{tab: survey info} as the required separations for each survey. We found eight unique stars, seven stars using FIRST and VLASS, six stars using FIRST and RACS-mid, and none using FIRST and RACS-low. This is likely because of higher noise in some RACS-low pointings and because the RACS-low catalogue convolved all of the images to the common resolution. There are 5 stars common to both VLASS and RACS-mid.
The positions of the stars in the various epochs and the separation between the source positions after proper-motion correction are shown in Table\,\ref{tab: star summary table} and the images are shown in Figure\,\ref{fig: star radio images}. We can see in the table the sources that are detected in more than one survey.
The flux densities of the sources from the radio surveys are shown in Table\,\ref{tab: star flux densities}, this includes sources that were not found in RACS-mid using the proper-motion method but do have a RACS-mid detection.  Here we will discuss each of the eight stars found using proper-motion searching.
 
\clearpage
\onecolumn

\begin{landscape}
\centering
\begin{small}
\begin{longtable}{P{1.4cm}A{1.4cm}rP{1.9cm}A{1.4cm}rrlP{2.4cm}A{1.4cm}rrrrr}
\caption[]{\label{tab: star summary table}Summary of the position (J2000 reference frame) information for each radio stellar source found using the proper-motion method. ``F'' 
 stands for FIRST. Survey B indicates the second radio survey used to find the source, either VLASS or RACS-mid. $D_{\mathrm{A,B}}$ is the separation in arcseconds between the position of the source in Survey A and the position of the source in Survey B, see Figure\,\ref{fig: Gaia radio pm source diagram}.}\\
 & \multicolumn{2}{c}{\textit{Gaia}}  & \multicolumn{4}{c}{FIRST} & \multicolumn{8}{c}{Survey B} \\
\hline
Name & Position &  Epoch &  Name & Position &  Epoch &  $D_{\mathrm{G_{F},F}}$ & Survey B & Name & Position &  Epoch &  $T_{\mathrm{F,B}}$ & $D_{\mathrm{G_{F,B}}}$ & $D_{\mathrm{F,B}}$ & $D_{\mathrm{G_{B},B}}$  \\
\hline
PM J15587$+$2351E  & 15:58:44.97 $+$23:51:17.87 & J2016.0 & FIRST J155845.0$+$235119 & 15:58:45.1 $+$23:51:19.5 & J1995.951 & 0.80\arcsec & RACS-mid & ASKAP J155844.94$+$235117.02 & 15:58:44.9 $+$23:51:17.0 & J2021.009 & 25.058\,yr & 2.93\arcsec & 3.31\arcsec & 0.62\arcsec \\ 
GS Leo  & 9:30:35.62 $+$10:36:06.04 & J2016.0 & FIRST J093035.8$+$103606 & 9:30:35.8 $+$10:36:06.3 & J2000.026 & 0.06\arcsec & VLASS & VLASS1QLCIR J093035.59$+$103606.1 & 9:30:35.6 $+$10:36:06.1 & J2017.752 & 17.726\,yr & 3.55\arcsec & 3.59\arcsec & 0.10\arcsec \\ 
& & & & & & & RACS-mid & ASKAP J93035.55$+$103605.31 & 9:30:35.6 $+$10:36:05.3 & J2021.044 & 21.018\,yr & 4.21\arcsec & 4.33\arcsec & 0.67\arcsec \\ 
sig CrB A & 16:14:40.51 $+$33:51:29.62& J2016.0 & FIRST J161440.9+33 & 16:14:41.0 $+$33:51:31.6 & J1994.471& 0.14\arcsec & VLASS& VLASS1QLCIR J161440.48$+$335129.5 & 7:43:18.6 $+$28:53:02.2 & J2017.762 & 23.291\,yr& 6.57\arcsec& 6.49\arcsec& 0.21\arcsec \\
 & &  &  &  & &  & RACS-mid& ASKAP J161440.40$+$335127.20 & 16:14:40.4 $+$33:51:27.2 & J2020.998& 26.527\,yr & 7.48\arcsec & 8.41\arcsec & 1.99\arcsec \\
sig Gem & 7:43:18.80 $+$28:52:56.96& J2016.0 & FIRST J074318.6$+$285302	 & 7:43:18.6 $+$28:53:02.2 & J1993.328& 0.64\arcsec & VLASS & VLASS1QLCIR J074318.83$+$285256.30 & 7:43:18.8 $+$28:52:56.3 & J2019.279& 25.951\,yr& 6.16\arcsec & 6.39\arcsec& 0.24\arcsec \\ & 
&  &  &  & &  & RACS-mid & ASKAP J74318.84$+$285255.10 & 7:43:18.8 $+$28:52:55.1 & J2021.011 & 27.683\,yr& 6.57\arcsec & 7.53\arcsec& 0.73\arcsec \\ 
BI Cet  & 1:22:50.17 $+$0:42:39.57 & J2016.0 & FIRST J012250.3$+$004243 & 1:22:50.3 $+$0:42:43.3 & J1998.774 & 0.63\arcsec & VLASS & VLASS1QLCIR J012250.16$+$004239.4 & 1:22:50.2 $+$0:42:39.4 & J2017.738 & 18.964\,yr & 5.03\arcsec & 4.62\arcsec & 0.30\arcsec \\ 
 & & & & & & & RACS-mid & ASKAP J12250.13$+$004238.77 & 1:22:50.1 $+$0:42:38.8 & J2021.029 & 22.255\,yr & 5.90\arcsec & 5.50\arcsec & 0.42\arcsec \\ 
 39 Cet  & 1:16:36.18 $-$2:30:02.35 & J2016.0 & FIRST J011636.2$-$023000	 & 1:16:36.3 $-$2:30:00.7 & J1998.172 & 0.49\arcsec & VLASS & VLASS1QLCIR J011636.16$-$023002.2 & 1:16:36.2 $-$2:30:02.3 & J2017.790 & 19.618\,yr & 2.33\arcsec & 2.48\arcsec & 0.17\arcsec \\ 
 & & & & & & & RACS-mid & 
ASKAP J11636.18$-$023003.74 & 1:16:36.2 $-$2:30:03.7 & J2020.991 & 
22.819\,yr & 2.71\arcsec & 3.49\arcsec & 1.18\arcsec \\ 
FK Com  & 13:30:46.74 $+$24:13:57.43 &
J2016.0 & FIRST J133046.8$+$241358 & 13:30:46.8 $+$24:13:58.4 & J1995.942 & 0.57\arcsec & VLASS & VLASS1QLCIR J133046.73$+$241357.4 & 13:30:46.7 $+$24:13:57.4 & J2017.751 & 21.809\,yr & 1.23\arcsec & 1.57\arcsec & 0.09\arcsec \\ 
BH CVn  & 13:34:47.92 $+$37:10:56.54 & J2016.0 & 	FIRST J133447.7$+$371056 & 13:34:47.8 $+$37:10:56.7 & J1994.560 & 0.10\arcsec & VLASS & VLASS1QLCIR J133447.95$+$371056.5 & 13:34:48.0 $+$37:10:56.5 & J2017.824 & 23.264\,yr & 2.00\arcsec & 2.13\arcsec & 0.22\arcsec \\
\end{longtable}
\end{small}
\end{landscape} 

\clearpage
\twocolumn

\begin{table}
\caption{Flux densities for the radio stellar sources found using proper motion. FK Com and BH CVn were not detected in RACS-mid using the proper-motion method as the RACS-mid position is $<3\arcsec$ from the FIRST position. However, we know that these sources are radio stars from FIRST--VLASS proper-motion matching. We have therefore included their RACS-mid flux densities in this table. Some sources are detected in RACS-mid more than once as they fall in the overlap between tiles. The FIRST survey catalogue does not include uncertainties on the peak flux density.}
    \centering
    \begin{tabular}{P{2.0cm}llA{1cm}rA{1.4cm}}
    Name & Survey & Freq (MHz) & Epoch & $F_{\mathrm{peak}}$ (mJy) \\
    \hline
    \hline
    PM\,J15587$+$2351E & FIRST & 1400.0 & J1995.95 & $1.3$ \\
     & RACS-MID & 1367.5 & J2021.01 & $1.6\pm0.2$ \\
    GS Leo & FIRST & 1400.0 & J2000.03 & $2.3$ \\
     & VLASS & 3000.0 & J2017.75 & $1.4\pm0.1$ \\
     & RACS-MID & 1367.5 & J2021.04 & $3.9\pm0.3$ \\
    sig CrB A & FIRST & 1400.0 & J1994.47 & $5.5$ \\
     & VLASS & 3000.0 & J2017.76 & $3.2\pm0.2$ \\
     & RACS-MID & 1367.5 & J2021.00 & $4.3\pm0.4$ \\
    sig Gem & FIRST & 1400.0 & J1993.33 & $2.0$ \\
     & VLASS & 3000.0 & J2019.28 & $54.2\pm0.1$ \\
     & RACS-MID & 1367.5 & J2021.01 & $2.6\pm0.3$ \\
    BI Cet & FIRST & 1400.0 & J1998.77 & $2.0$ \\
     & VLASS & 3000.0 & J2017.74 & $3.4\pm0.1$ \\
     & RACS-MID & 1367.5 & J2021.03 & $1.4\pm0.2$ \\
    39 Cet & FIRST & 1400.0 & J1998.17 & $2.2$ \\
     & VLASS & 3000.0 & J2017.79 & $2.7\pm0.2$ \\
     & RACS-MID & 1367.5 & J2020.99 & $3.0\pm0.5$ \\
     & RACS-MID & 1367.5 & J2020.99 & $2.4\pm0.3$ \\
    FK Com & FIRST & 1400.0 & J1995.94 & $1.8$ \\
     & VLASS & 3000.0 & J2017.75 & $8.0\pm0.1$ \\
     & RACS-MID & 1367.5 & J2020.99 & $2.7\pm0.3$ \\
     & RACS-MID & 1367.5 & J2021.00 & $2.0\pm0.3$ \\
    BH CVn & FIRST & 1400.0 & J1994.56 & $4.0$ \\
     & VLASS & 3000.0 & J2017.82 & $9.5\pm0.1$ \\
     & RACS-MID & 1367.5 & J2020.98 & $13.1\pm0.8$ \\
    \hline
    \end{tabular}
    \label{tab: star flux densities}
 \end{table}

 \begin{figure*}
    \caption{\label{fig: star radio images} Radio images of the stellar sources found using radio proper-motion.
    The cross-hairs indicate the \gaia\,DR3 proper-motion corrected position corrected to the epoch of the radio image. The circles indicate the radio position of the source and the radius is the uncertainty on the radio position: FIRST, cyan, $1\arcsec$; VLASS, magenta, $0.5\arcsec$; and RACS-mid, yellow, $2\arcsec$; . The grey scale is not the same for every panel. PM\,J15587$+$2351E is not detected by VLASS. Both FK\,Com and BH\,CVn were only found using the proper-motion method with FIRST and VLASS, they were not found using FIRST and RACS-mid as the separation between the FIRST position and the RACS-mid position is $<3\arcsec$. However, both sources are detected by RACS-mid, as we can see in these plots.}
        \begin{subfigure}{\textwidth}
            \includegraphics[width=0.95\textwidth]{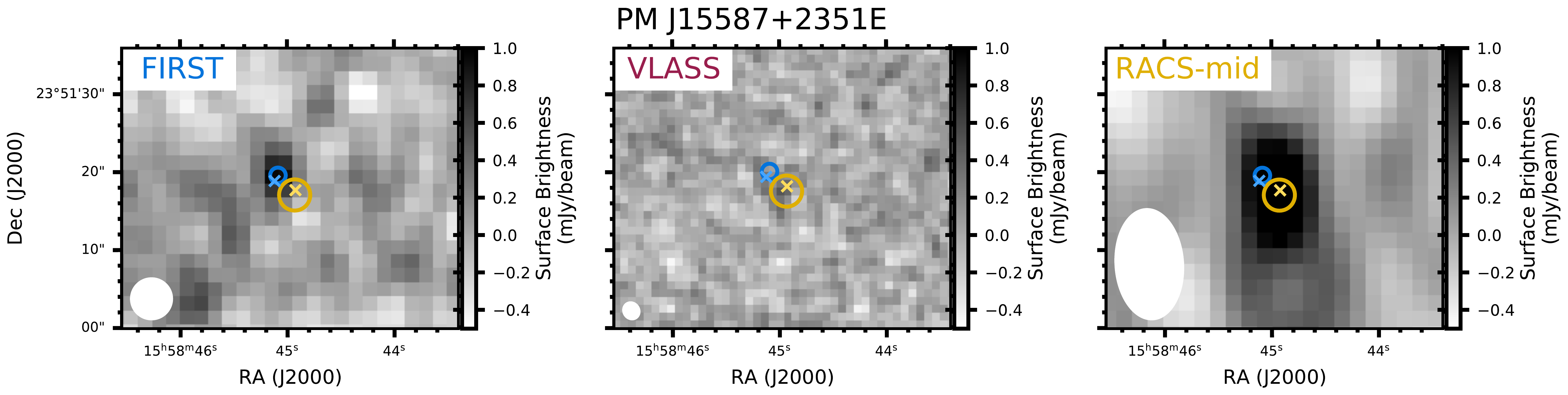}
        \end{subfigure}
        \begin{subfigure}{\textwidth}
            \includegraphics[width=0.95\textwidth]{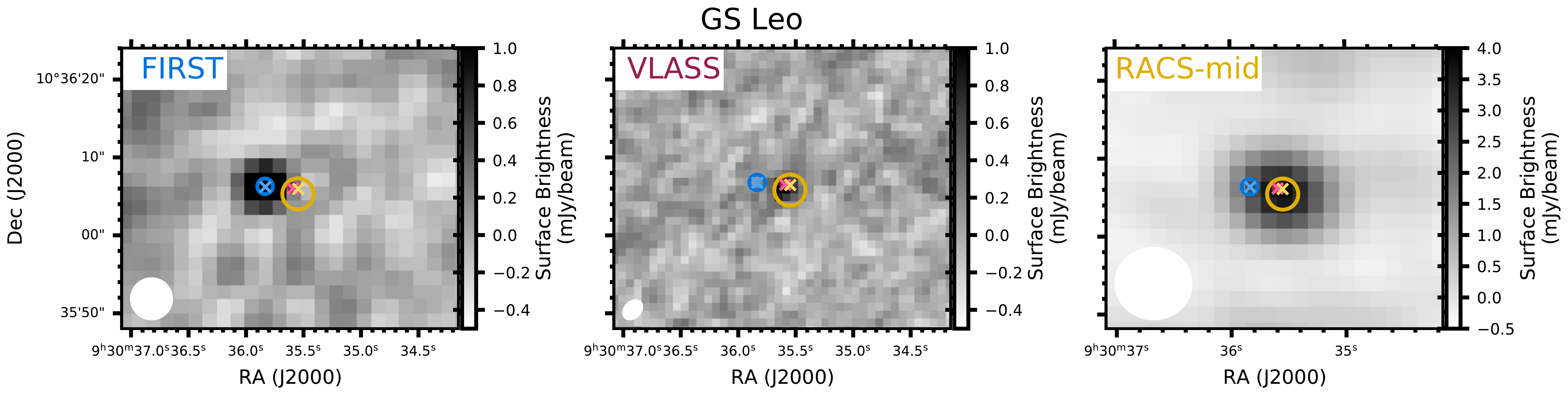}
        \end{subfigure}
        \begin{subfigure}{\textwidth}
            \includegraphics[width=0.95\textwidth]{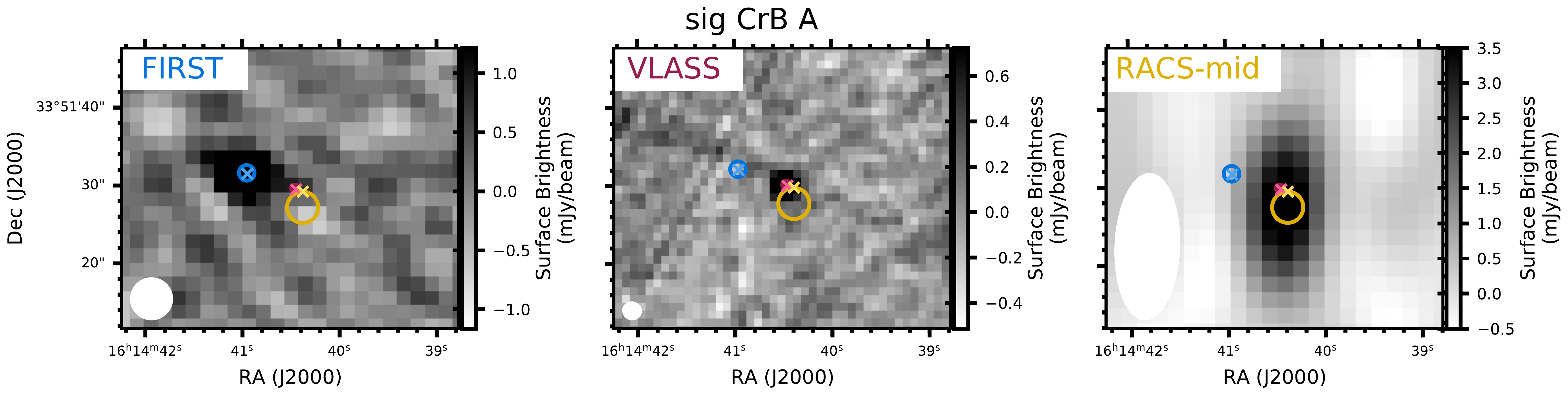}
        \end{subfigure}
        \begin{subfigure}{\textwidth}
            \includegraphics[width=0.95\textwidth]{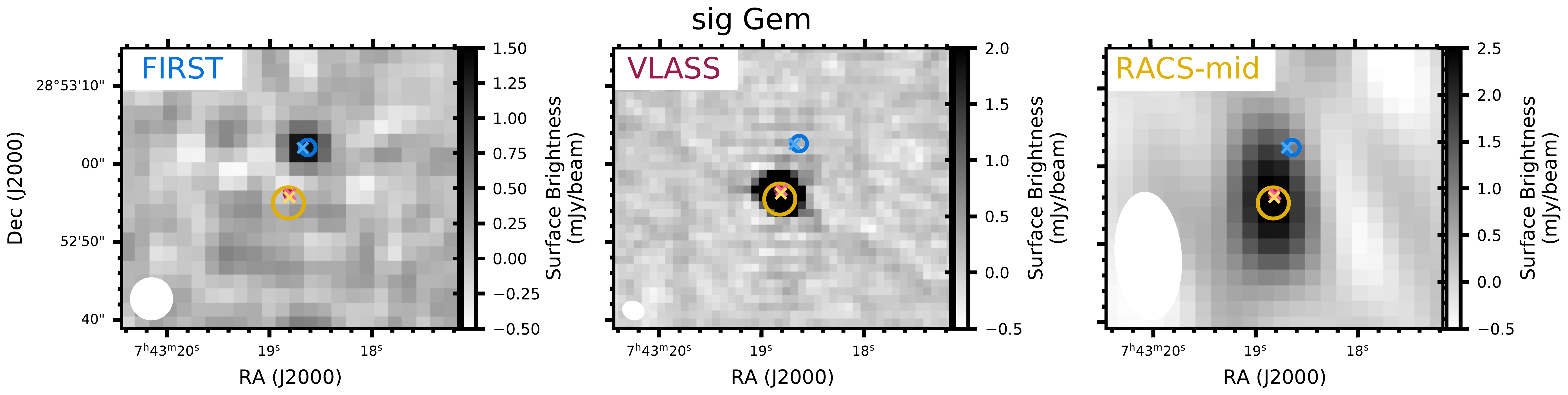}
        \end{subfigure}
        \begin{subfigure}{\textwidth}
            \includegraphics[width=0.95\textwidth]{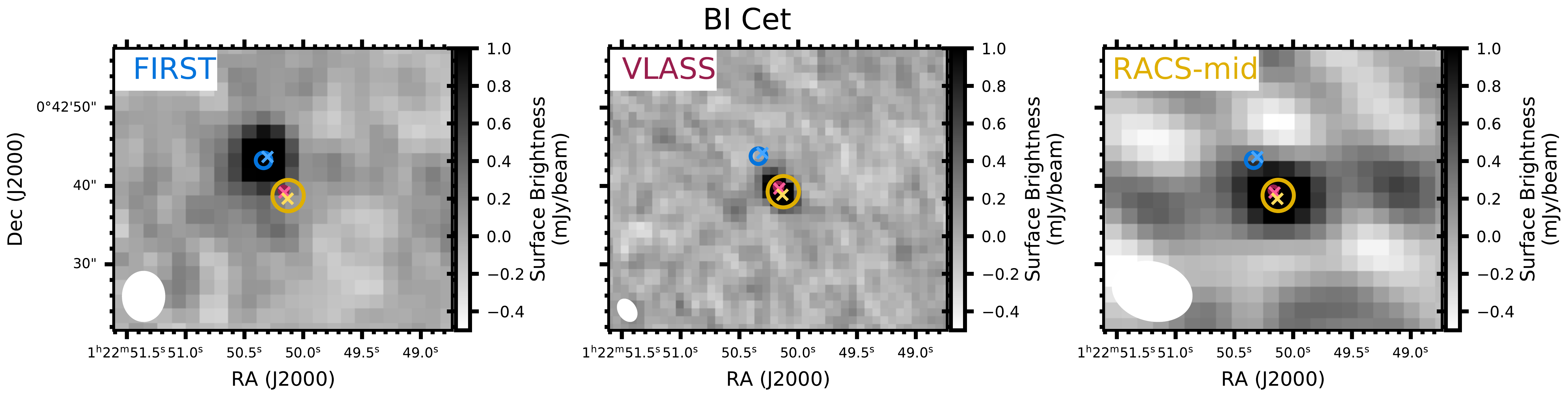}
        \end{subfigure}
\end{figure*}

\begin{figure*}
    \ContinuedFloat
    \captionsetup{list=off,format=cont}
    \caption{}
        \begin{subfigure}{\textwidth}
            \includegraphics[width=0.95\textwidth]{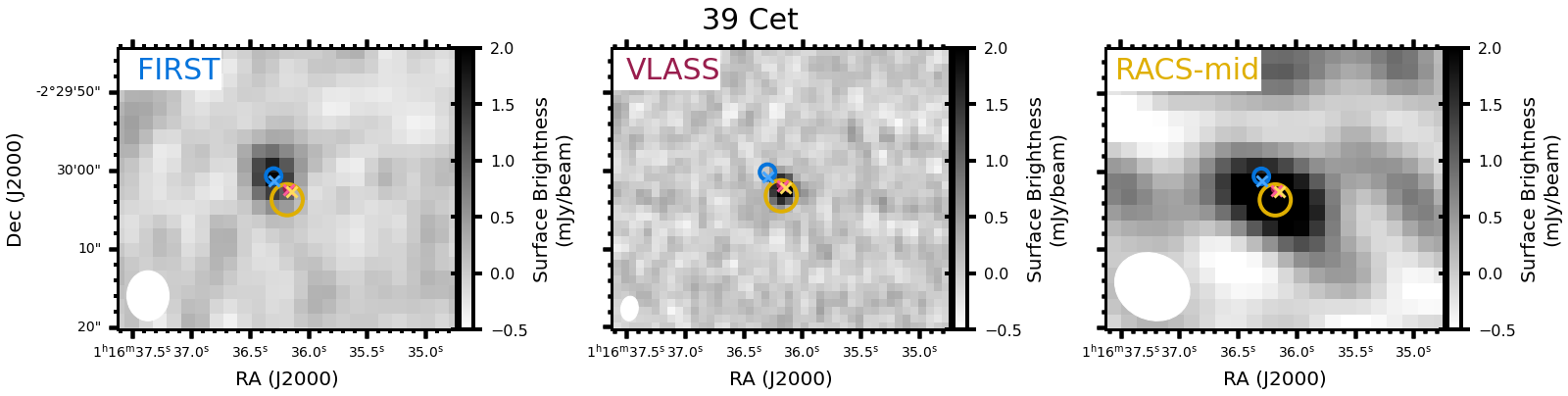}
        \end{subfigure}
        \begin{subfigure}{\textwidth}
            \includegraphics[width=0.95\textwidth]{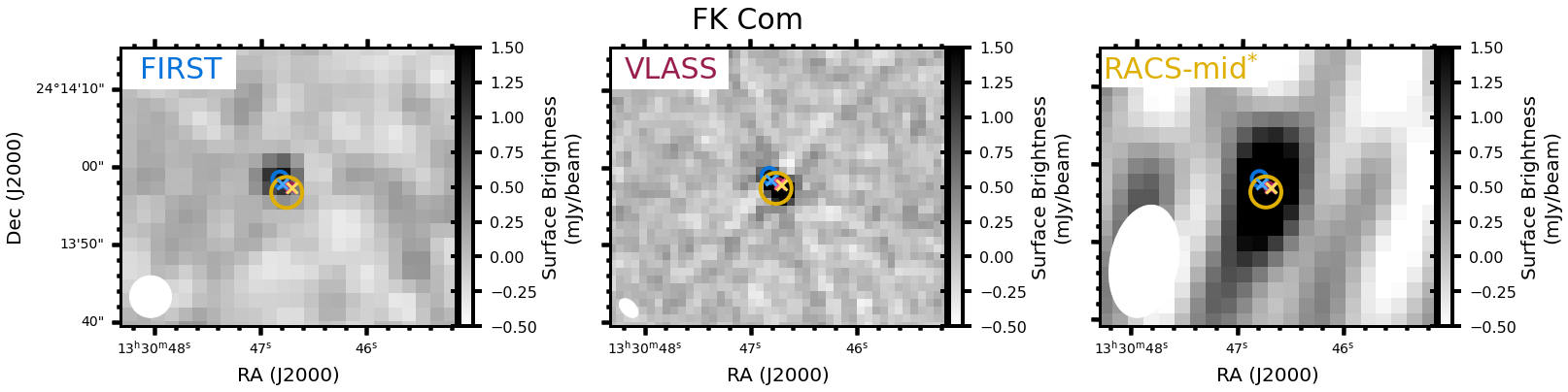}
        \end{subfigure}
        \begin{subfigure}{\textwidth}
            \includegraphics[width=0.95\textwidth]{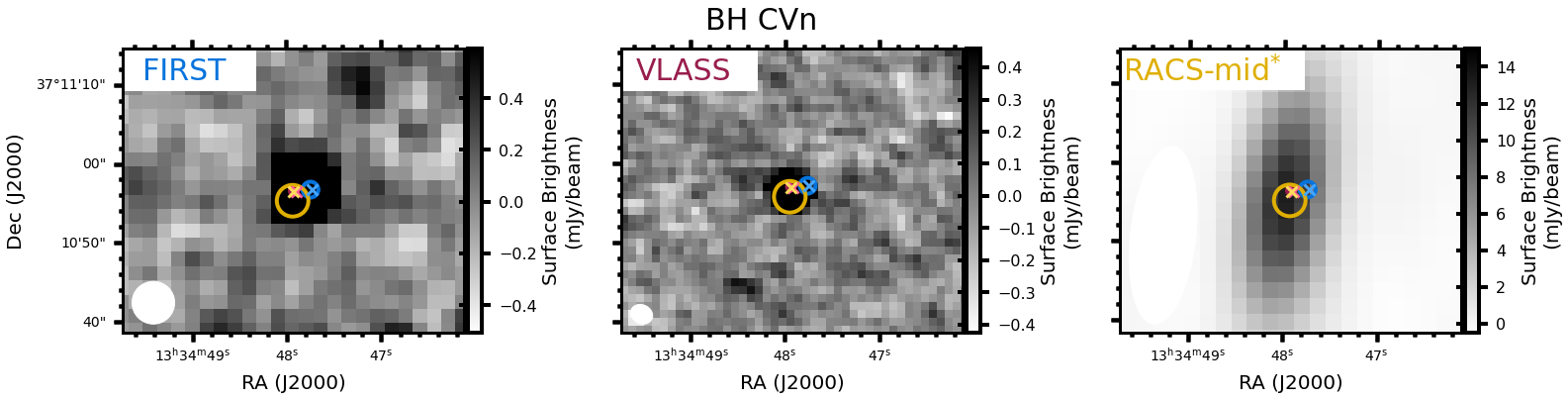}
        \end{subfigure}
\end{figure*}

\textbf{PM\,J15587$+$2351E} has not previously been identified as a radio star. It is an M5e D type M dwarf star \citep[e.g.][]{2016MNRAS.457.2192C,2019ApJ...877...60B} and is $35.4\pm0.1$\,pc away \citep{2021AJ....161..147B}.

\textbf{GS Leo} has not previously been identified as a radio star. It is a known X-ray star and is known to be variable. GS Leo is a tight 3.5\,d G9V+K4 binary in a wide binary with a K0 star \citep[e.g.][]{2012AN....333..663S,2017ARep...61...80S} that is $35.37\pm0.04$\,pc away \citep{2021AJ....161..147B}.

\textbf{$\sigma$ Coronae Borealis (sig CrB) A} has previously been detected in the radio using VLBI to discern the stellar radio emission from a nearby QSO \citep[][]{1992A&A...258..112L}. Similar to our method, they used the proper-motion to determine that the radio emission is moving across the sky. $\sigma$ CrB A has recently been detected at 144\,MHz in both Stokes I and Stokes V using LOFAR \citep[][]{2021A&A...654A..21T}. It is a spectroscopic, RS CVn binary and is $22.68^{+0.03}_{-0.02}$\,pc away \citep{2021AJ....161..147B}.

\textbf{$\sigma$ Gemini (sig Gem)} is a known FIRST radio star and it has been detected by LOFAR in Stokes I at 144\,MHz \citep[][]{2022ApJ...926L..30V}. \citet{1977AJ.....82..989S} made the first tentative radio detection. It is an RS CVn type binary with a K1III component \citep[e.g.][]{2019A&A...623A..72K} and is $36.9\pm0.5$\,pc away \citep{2021AJ....161..147B}.

\textbf{BI Ceti (BI Cet)} is a known FIRST radio star. It was first detected in the radio by \citet{1986AJ.....91.1229D} using the VLA. BI Cet is an RS CVn binary consisting of a G6IV/V star and a G6V star \citep{2015ARep...59..937K} $62.12^{+0.07}_{-0.1}$\,pc away \citep{2021AJ....161..147B}.

\textbf{39 Ceti (39 Cet)} is a known FIRST radio star. It was first detected in the radio by \citet{1982BAAS...14Q.982S} using the VLA. It is an RS CVn type binary consisting of a G5III star and a DA2.8 white dwarf \citep[e.g.][]{2015AJ....150...88L} $74.6\pm0.5$\,pc away \citep{2021AJ....161..147B}.

\textbf{FK Comae Berenices (FK Com)} is a known FIRST radio star. It has been detected in both Stokes I and Stokes V at 144\,MHz with LOFAR \citep[][]{2021A&A...654A..21T}. FK Com is an FK Com type G star (a single star that was a binary, but merged) $222.8^{+1.4}_{-1.3}$\,pc away \citep{2021AJ....161..147B}.

\textbf{BH Canum Venaticorum (BH CVn)} is a known FIRST and NVSS radio star and it has also been detected at 144\,MHz in both Stokes I and V by LOFAR \citep[][]{2021A&A...654A..21T,2022ApJ...926L..30V}. This source was found by \citet{2021MNRAS.502.5438P} in their search for circularly polarised sources with ASKAP. BH CVn is an RS CVn type binary and is $46.8\pm0.2$\,pc away \citep{2021AJ....161..147B}.

\section{Candidate variable stellar radio sources}
\label{sec: demo unmatched}

\begin{figure}
\caption{Diagram illustrating two \gaia\,sources which would be considered candidate radio variable stellar sources. In both cases (a) and (b) the optical source does match a radio source in survey A but does not match a source in survey B.
}
\includegraphics[width=\columnwidth]{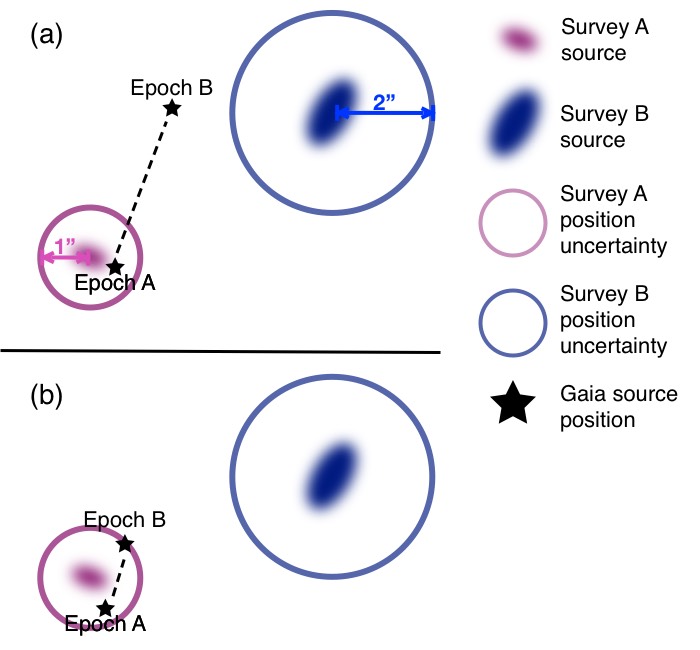}
\label{fig: radio proper motion variable}
\end{figure}

In Section\,\ref{sec: method} we described our steps for proper-motion matching. In step\,\ref{step: xmatch B Gaia} we discard the \gaia\,sources and corresponding survey A sources that do not match a survey B source. However, if survey B is equally sensitive or more sensitive than survey A it is interesting to explore the survey A and \gaia\,matches that do not correspond to a survey B source. This is because it means that a radio source was detected in an earlier epoch but was not detected in a later epoch even though the later epoch is from a more sensitive survey. A diagram demonstrating sources that would be considered variables in this way is shown in Figure\,\ref{fig: radio proper motion variable}. These sources are likely radio variables, and may indicate that the initial radio detection was from e.g. a stellar flare. Due to the filtering we do to search for proper-motion stars, this is not a complete search for radio variables. We will only find radio variables that match an optical source, which means we will miss those radio variables where the survey A source does not have an optical match. We also only match to optical sources with a proper-motion measurement in \gaia, further reducing the sample.

The steps we used to find candidate variable stars are as follows:
\begin{enumerate}
    \item \label{step: cut Gaia pm var} {Discard \gaia\,sources that have no measured proper-motion magnitude}
    \item \label{step: resolved cut var} Discard sources in survey A where $F_{\mathrm{int}}/F_{\mathrm{peak}}\geq 1.5$. Do \textbf{not} do this for survey B.
    \item \label{step: discard A B var} Keep any survey A sources that are separated by $>a\arcsec+b\arcsec$ from all survey B sources and keep any survey B sources that are separated by $>a\arcsec+b\arcsec$ from all survey A sources
    \item \label{step: xmatch A Gaia var} Proper-motion correct the \gaia\,source positions to the survey A epoch and cross-match the source positions. The sources are considered a match if the separation is $<a\arcsec$. Discard those \gaia\,and survey A sources that do not match.
    \item \label{step: xmatch B Gaia var} Proper-motion correct the remaining \gaia\,source positions to the survey B epoch and cross-match the source positions. The sources are considered a match if the separation is $<b\arcsec$. Discard those \gaia\, and survey B sources that \textbf{do} match.
    \item \label{step: parallax cut var} Discard \gaia\,sources (and the corresponding FIRST matches) that have a {parallax over error $<5$}.
\end{enumerate}
We perform these steps with FIRST as survey A and treat each of VLASS, RACS-low and RACS-mid separately as survey B. However, there are sources in common between the different surveys. At this point, we combine the three separate FIRST-VLASS, FIRST-RACS-low, FIRST-RACS-mid catalogues into one catalogue. We are then working with the combined catalogue for the following steps:
\begin{enumerate}
\setcounter{enumi}{6}
    \item \label{step: racs-mid only var} Discard sources where $D_{\mathrm{FIRST},\mathrm{RACS-mid}}<3\arcsec$
    \item \label{step: racs-mid only covered var} Discard sources outside the RACS-mid FoV, i.e. sources with declination $>45^{\circ}$
    \item \label{step: vlass cut var} Discard sources where $D_{\mathrm{FIRST},\mathrm{VLASS}}<D_{\mathrm{G_{VLASS}},\mathrm{VLASS}}$
\end{enumerate}
In step (\ref{step: resolved cut var}) we do not remove resolved sources from survey B as faint point sources in survey A might be revealed as extended sources in survey B.
In step (\ref{step: xmatch B Gaia var}) we remove the sources that do have a proper-motion match, as the sources with a proper-motion match are the stars presented in Section\,\ref{sec: demo results}.
In step (\ref{step: parallax cut var}) we use the \gaia\,parallax over error, the parallax value divided by the uncertainty, to reduce the number of extra-galactic sources in our sample. {This is because the parallax over error represents the signal to noise of the parallax measurement, and extra-galactic sources typically have {lower uncertainty-normalised parallax measurements as they are fainter}}.
We perform steps (\ref{step: cut Gaia pm var}) to (\ref{step: parallax cut var}) on each of VLASS, RACS-low and RACS-mid, with FIRST as survey A. However, RACS-mid (1367\,MHz) is the closest in frequency to FIRST (1400\,MHz), so we find these sources to be the strongest candidates. This is why we include step (\ref{step: racs-mid only var}).
We find that there are some sources where the VLASS-FIRST separation ($D_{\mathrm{FIRST},\mathrm{VLASS}}$) is less than the VLASS-proper-motion corrected \gaia\,position separation ($D_{\mathrm{G_{VLASS}},\mathrm{VLASS}}$).
This means that the optical source is less likely to be responsible for the radio emission. We remove such sources in step (\ref{step: vlass cut var}).

After performing steps (\ref{step: cut Gaia pm var}) to (\ref{step: xmatch B Gaia var}) we are left with FIRST sources that are matched to \gaia\,proper-motion sources, but do not have a match in one of VLASS, RACS-low or RACS-mid either at the FIRST position or at the \gaia\,proper-motion corrected position. At this point, we find 4638, 4748, and 2768 candidate variable radio sources when we compare FIRST to VLASS, RACS-low, and RACS-mid respectively\footnote{The variable source candidate catalogue at this point is available in the supplementary material}. After these steps, many of these sources are likely to still be extra-galactic sources. We do not investigate most of these sources further.

After applying the parallax cut (step (\ref{step: parallax cut var})), we are left with 76, 88, and 115 candidate variable radio stellar sources when we compare FIRST to VLASS, RACS-low, and RACS-mid respectively. It is at this point that we combine the individual VLASS, RACS-low, and RACS-mid candidates into one catalogue of 156 total unique candidate variable radio stellar sources. After performing steps (\ref{step: racs-mid only var}) to (\ref{step: vlass cut var}) we have 62 remaining candidate variable radio stellar sources. Finally, we manually check the FIRST and VLASS images and remove any sources that are extended or appear to be artefacts. This eliminates 8 sources, leaving 54 sources in our final set of candidate variable radio stellar sources.
We present the \gaia\,positions and the separations between the FIRST position and nearest VLASS, RACS-low and RACS-mid sources in Table\,\ref{tab: variable star cands}.

\clearpage
\onecolumn

\begin{landscape}
\centering
\begin{small}
\begin{longtable}{P{3.8cm}A{1.4cm}rA{0.9cm}A{0.9cm}rA{0.9cm}A{0.9cm}rA{0.9cm}A{0.9cm}rA{0.9cm}A{0.9cm}}
\caption[]{\label{tab: variable star cands}{Details of the 54 candidate variable radio stellar sources.
We include both the optical/Simbad name (or \gaia\,DR3 designation where a Simbad name is not available) and the FIRST name for each source.
In the columns where we give the separations between source positions, ``F'' is for FIRST, ``RL'' is for RACS-low, ``RM'' is for RACS-mid and ``V'' is for VLASS. The FIRST survey catalogue does not include uncertainties on the peak flux density and the typical RMS noise values for FIRST, RACS-low, RACS-mid and VLASS are shown in Table\,\ref{tab: survey info}. A machine-readable version of this table is available in the supplementary material.
}}\\
 &  & \multicolumn{3}{c}{FIRST} & \multicolumn{3}{c}{VLASS} & \multicolumn{3}{c}{RACS-low} & \multicolumn{3}{c}{RACS-mid} \\
\hline
Name & Gaia J2016 & Epoch & Gaia sep (\arcsec) & Flux density (mJy) & Epoch & Gaia sep (\arcsec) & FIRST sep (\arcsec) & Epoch & Gaia sep (\arcsec) & FIRST sep (\arcsec) & Epoch & Gaia sep (\arcsec) & FIRST sep (\arcsec) \\ 
\hline
\hline
2MASS J07500030$+$3458579 FIRST\,J075000.2$+$345858 & 7:50:00.3 $+$34:58:57.7 & J1994.5 & 0.78 & 1.82 & J2019.4 & 0.69 & 186.33 &  &  &  & J2021.0 & 0.73 & 3.41 \\
2MASS J08293480$+$0858099 FIRST\,J082934.8$+$085809 & 8:29:34.8 $+$8:58:10.2 & J2000.1 & 0.65 & 1.41 & J2017.8 & 0.52 & 68.03 & J2019.3 & 0.57 & 69.59 & J2021.0 & 0.62 & 68.04 \\
2MASS J09244488$+$0019097 FIRST\,J092444.8$+$001908 & 9:24:44.9 $+$0:19:09.7 & J1998.6 & 0.91 & 1.98 & J2018.0 & 0.08 & 338.12 & J2019.3 & 0.08 & 178.17 & J2021.0 & 0.09 & 178.22 \\
2MASS J09420757$+$0334344 FIRST\,J094207.5$+$033434 & 9:42:07.6 $+$3:34:34.4 & J1998.5 & 0.47 & 1.15 & J2018.0 & 0.10 & 14.30 & J2020.3 & 0.11 & 9.46 & J2021.0 & 0.12 & 13.28 \\
2MASS J09582770$+$2847572 FIRST\,J095827.7$+$284757 & 9:58:27.8 $+$28:47:57.2 & J1993.3 & 0.87 & 1.06 & J2019.3 & 0.95 & 324.92 & J2020.3 & 0.99 & 94.92 & J2021.0 & 1.01 & 79.33 \\
2MASS J10014486$+$2756455 FIRST\,J100144.7$+$275645 & 10:01:44.8 $+$27:56:45.2 & J1995.8 & 0.93 & 1.03 & J2019.3 & 0.81 & 98.80 & J2020.3 & 0.84 & 98.26 & J2021.0 & 0.86 & 98.89 \\
2MASS J14333139$+$3417472 FIRST\,J143331.5$+$341747 & 14:33:31.3 $+$34:17:46.8 & J1994.5 & 0.87 & 1.23 & J2017.9 & 2.28 & 249.63 &  &  &  & J2021.0 & 2.58 & 6.88 \\
2MASS J15085996$+$2714307 FIRST\,J150859.9$+$271430 & 15:09:00.0 $+$27:14:30.5 & J1995.9 & 0.57 & 1.06 & J2017.8 & 0.36 & 397.11 & J2020.3 & 0.40 & 31.03 & J2021.0 & 0.41 & 181.18 \\
2MASS J15215160$+$4246246 FIRST\,J152151.6$+$424624 & 15:21:51.6 $+$42:46:24.5 & J1996.0 & 0.24 & 2.4 & J2019.2 & 0.20 & 0.61 &  &  &  & J2021.0 & 0.21 & 3.74 \\
2MASS J16234398$+$1302124 FIRST\,J162343.9$+$130211 & 16:23:44.0 $+$13:02:11.9 & J2000.0 & 0.95 & 1.08 & J2019.3 & 0.65 & 188.54 & J2019.3 & 0.65 & 188.55 & J2021.1 & 0.71 & 60.35 \\
2MASS J20485716$-$0053473 FIRST\,J204857.2$-$005348 & 20:48:57.3 $-$0:53:48.5 & J2011.3 & 0.15 & 1.94 & J2018.0 & 0.67 & 221.63 & J2019.3 & 0.80 & 221.93 & J2021.0 & 0.97 & 220.96 \\
BD$+$09  4984B FIRST\,J220611.8$+$100528 & 22:06:11.8 $+$10:05:28.6 & J2011.2 & 0.57 & 2.22 & J2017.7 & 0.04 & 271.77 & J2019.3 & 0.05 & 367.67 & J2021.6 & 0.07 & 271.80 \\
BD$-$08  6022 FIRST\,J230553.0$-$074548 & 23:05:53.0 $-$7:45:48.9 & J2011.3 & 0.18 & 3.74 & J2017.9 & 0.23 & 74.80 & J2019.3 & 0.27 & 52.94 & J2021.0 & 0.33 & 53.73 \\
LP  521$-$15 FIRST\,J224656.1$+$143715 & 22:46:56.3 $+$14:37:15.3 & J2011.3 & 0.80 & 1.67 & J2017.8 & 2.23 & 377.34 & J2020.5 & 3.17 & 376.09 & J2021.0 & 3.35 & 377.14 \\
Gaia DR3 1158730503710098688 FIRST\,J144724.8$+$043700 & 14:47:24.9 $+$4:37:00.4 & J2000.7 & 0.57 & 1.06 & J2019.2 & 0.05 & 143.84 & J2019.3 & 0.05 & 145.28 & J2021.1 & 0.05 & 145.65 \\
Gaia DR3 1360691101602998784 FIRST\,J172747.2$+$440031 & 17:27:47.2 $+$44:00:31.8 & J1995.6 & 0.59 & 1.38 & J2019.3 & 0.14 & 55.28 &  &  &  & J2021.0 & 0.15 & 50.71 \\
Gaia DR3 1732181789609582208 FIRST\,J211059.8$+$042632 & 21:11:00.0 $+$4:26:33.0 & J2009.2 & 0.77 & 1.55 & J2017.8 & 0.06 & 161.75 & J2019.3 & 0.07 & 160.18 & J2021.1 & 0.08 & 161.23 \\
Gaia DR3 2377310715263856512 FIRST\,J003921.0$-$111037 & 0:39:21.1 $-$11:10:37.4 & J1997.4 & 0.61 & 1.08 & J2017.9 & 0.25 & 120.27 & J2020.2 & 0.28 & 42.83 & J2021.0 & 0.29 & 37.07 \\
Gaia DR3 2467183490048356864 FIRST\,J014456.9$-$071309 & 1:44:56.9 $-$7:13:09.1 & J2009.3 & 0.98 & 1.01 & J2017.9 & 0.38 & 49.55 & J2019.3 & 0.44 & 49.79 & J2021.0 & 0.51 & 49.96 \\
Gaia DR3 2490713524213973248 FIRST\,J020618.5$-$055910 & 2:06:18.5 $-$5:59:10.3 & J1997.4 & 0.61 & 1.05 & J2017.9 & 0.23 & 575.65 & J2020.2 & 0.25 & 131.36 & J2021.0 & 0.26 & 131.28 \\
Gaia DR3 2633166727449471360 FIRST\,J233453.9$-$043809 & 23:34:53.9 $-$4:38:09.3 & J2011.2 & 0.31 & 2.67 & J2017.9 & 0.06 & 88.05 & J2019.3 & 0.08 & 89.06 & J2021.0 & 0.09 & 58.60 \\
Gaia DR3 2715431259726742784 FIRST\,J230214.6$+$104205 & 23:02:14.7 $+$10:42:05.9 & J2011.3 & 0.68 & 1.69 & J2017.8 & 0.33 & 99.21 & J2019.3 & 0.40 & 98.71 & J2021.6 & 0.52 & 98.74 \\
Gaia DR3 2776753256590751104 FIRST\,J005348.1$+$124519 & 0:53:48.1 $+$12:45:18.9 & J2011.2 & 0.88 & 1.25 & J2017.8 & 0.16 & 282.10 & J2019.3 & 0.20 & 292.79 & J2021.0 & 0.24 & 114.92 \\
 &  & \multicolumn{3}{c}{FIRST} & \multicolumn{3}{c}{VLASS} & \multicolumn{3}{c}{RACS-low} & \multicolumn{3}{c}{RACS-mid} \\
\hline
Name & Gaia J2016 & Epoch & Gaia sep (\arcsec) & Flux density (mJy) & Epoch & Gaia sep (\arcsec) & FIRST sep (\arcsec) & Epoch & Gaia sep (\arcsec) & FIRST sep (\arcsec) & Epoch & Gaia sep (\arcsec) & FIRST sep (\arcsec) \\ 
\hline
\hline
Gaia DR3 3147019745776728448 FIRST\,J075700.2$+$091955 & 7:57:00.2 $+$9:19:55.0 & J2000.1 & 0.69 & 1.25 & J2017.9 & 0.06 & 2.03 & J2019.3 & 0.07 & 3.81 & J2021.0 & 0.07 & 165.97 \\
Gaia DR3 3592258045911461376 FIRST\,J075700.2$+$091955 & 11:42:12.2 $-$7:53:59.4 & J2001.4 & 0.12 & 1.0 & J2019.4 & 0.32 & 173.30 & J2019.3 & 0.32 & 175.22 & J2021.0 & 0.35 & 174.17 \\
Gaia DR3 6910163884580689280 FIRST\,J205141.4$-$063340 & 20:51:41.4 $-$6:33:40.6 & J2011.3 & 0.41 & 1.03 & J2019.3 & 0.07 & 86.33 & J2019.3 & 0.07 & 85.85 & J2021.0 & 0.08 & 85.17 \\
SDSS J161007.07$+$394132.8 FIRST\,J161007.0$+$394132 & 16:10:07.1 $+$39:41:32.7 & J1994.6 & 0.48 & 1.13 & J2017.8 & 0.17 & 143.76 &  &  &  & J2021.0 & 0.19 & 231.72 \\
StKM 1$-$1155 FIRST\,J142555.9$+$141210 & 14:25:55.9 $+$14:12:09.6 & J2000.0 & 0.51 & 1.73 & J2019.3 & 1.29 & 502.20 & J2020.3 & 1.36 & 2.16 & J2021.1 & 1.41 & 453.37 \\
TYC 2503$-$1270$-$1 FIRST\,J100502.4$+$301824 & 10:05:02.4 $+$30:18:24.0 & J1993.3 & 0.53 & 1.35 & J2019.3 & 0.99 & 315.37 &  &  &  & J2021.0 & 1.05 & 3.91 \\
UCAC4 431$-$063012 FIRST\,J155042.6$-$035846 & 15:50:42.6 $-$3:58:46.6 & J1999.0 & 0.96 & 1.68 & J2019.3 & 0.52 & 170.52 & J2019.3 & 0.52 & 169.55 & J2022.4 & 0.60 & 170.01 \\
V* FF Aqr FIRST\,J220036.4$-$024427 & 22:00:36.5 $-$2:44:27.0 & J2011.2 & 0.26 & 2.39 & J2017.7 & 0.22 & 0.53 & J2019.3 & 0.27 & 235.86 & J2021.0 & 0.32 & 113.26 \\
V* IN Leo FIRST\,J103959.0$+$132722 & 10:39:59.0 $+$13:27:21.6 & J2000.0 & 0.28 & 1.95 & J2018.0 & 0.47 & 0.60 & J2019.3 & 0.50 & 102.61 & J2021.0 & 0.55 & 102.02 \\
V* V436 Ser FIRST\,J152346.1$-$004424 & 15:23:46.1 $-$0:44:24.8 & J1998.6 & 0.50 & 2.63 & J2019.3 & 0.28 & 209.57 & J2019.3 & 0.28 & 209.95 & J2022.4 & 0.33 & 209.51 \\
Wolf  424 A FIRST\,J123317.3$+$090115 & 12:33:15.5 $+$9:01:19.5 & J2000.1 & 0.23 & 1.41 & J2019.3 & 34.78 & 200.94 & J2020.3 & 36.54 & 282.36 & J2021.0 & 37.80 & 276.49 \\
Wolf  424 B FIRST\,J123317.3$+$090115 & 12:33:15.5 $+$9:01:18.6 & J2000.1 & 0.80 & 1.41 & J2019.3 & 33.12 & 200.94 & J2020.3 & 34.79 & 282.36 & J2021.0 & 36.00 & 276.49 \\
$[$UBW2009$]$ 21 FIRST\,J090624.3$+$001537 & 9:06:24.4 $+$0:15:38.0 & J1998.6 & 0.97 & 1.04 & J2017.7 & 0.24 & 265.78 & J2019.3 & 0.26 & 265.39 & J2021.0 & 0.29 & 265.48 \\
Gaia DR3 1157257742244515584 FIRST\,J150822.1$+$061436 & 15:08:22.2 $+$6:14:37.5 & J2000.1 & 0.73 & 1.02 & J2019.2 & 0.08 & 257.55 & J2020.3 & 0.08 & 258.67 & J2021.1 & 0.08 & 123.80 \\
Gaia DR3 1305574679646964352 FIRST\,J162716.0$+$275658 & 16:27:16.0 $+$27:56:57.8 & J1995.8 & 1.00 & 1.19 & J2017.9 & 0.22 & 15.74 & J2020.3 & 0.24 & 16.01 & J2021.0 & 0.25 & 14.46 \\
2MASS J03065628$+$0044316 FIRST\,J030656.2$+$004431 & 3:06:56.3 $+$0:44:31.6 & J1997.4 & 0.52 & 0.97 & J2017.9 & 0.04 & 279.99 & J2019.3 & 0.04 & 281.11 & J2021.0 & 0.05 & 280.04 \\
2MASS J07415981$+$2331589 FIRST\,J074159.8$+$233158 & 7:41:59.8 $+$23:31:58.9 & J1996.0 & 0.81 & 1.29 & J2019.3 & 0.08 & 183.03 & J2019.3 & 0.08 & 192.39 & J2021.0 & 0.08 & 192.81 \\
2MASS J16115068$+$4344126 FIRST\,J161150.6$+$434412 & 16:11:50.7 $+$43:44:12.7 & J1994.6 & 0.23 & 1.2 & J2019.3 & 0.26 & 61.03 &  &  &  & J2021.0 & 0.28 & 46.49 \\
2MASS J16235772$+$2350113 FIRST\,J162357.7$+$235010 & 16:23:57.7 $+$23:50:11.2 & J1996.0 & 0.63 & 1.88 & J2017.7 & 0.23 & 120.75 & J2020.3 & 0.26 & 126.80 & J2021.0 & 0.27 & 110.83 \\
FBQS J0748$+$3709 FIRST\,J074809.7$+$370926 & 7:48:09.8 $+$37:09:26.1 & J1994.6 & 0.61 & 1.05 & J2019.3 & 0.10 & 96.21 &  &  &  & J2021.0 & 0.11 & 98.10 \\
FBQS J0754$+$3937 FIRST\,J075413.7$+$393720 & 7:54:13.8 $+$39:37:20.0 & J1994.6 & 0.67 & 1.25 & J2019.3 & 0.11 & 214.60 &  &  &  & J2021.0 & 0.12 & 6.61 \\
FBQS J1216$+$3020 FIRST\,J121624.2$+$302042 & 12:16:24.2 $+$30:20:42.3 & J1993.3 & 0.60 & 1.06 & J2017.9 & 0.45 & 37.68 &  &  &  & J2021.0 & 0.50 & 38.52 \\
FBQS J1421$+$3319 FIRST\,J142142.1$+$331935 & 14:21:42.1 $+$33:19:35.7 & J1995.0 & 0.99 & 1.12 & J2017.9 & 0.44 & 215.94 &  &  &  & J2021.0 & 0.50 & 4.29 \\
FBQS J1704$+$2931 FIRST\,J170411.6$+$293153 & 17:04:11.6 $+$29:31:52.9 & J1994.7 & 0.61 & 1.13 & J2017.9 & 0.24 & 168.86 &  &  &  & J2021.0 & 0.27 & 180.45 \\
FBQS J1707$+$3802 FIRST\,J170718.5$+$380204 & 17:07:18.5 $+$38:02:04.4 & J1994.6 & 0.52 & 1.04 & J2017.9 & 0.16 & 65.35 &  &  &  & J2021.0 & 0.18 & 63.79 \\
 &  & \multicolumn{3}{c}{FIRST} & \multicolumn{3}{c}{VLASS} & \multicolumn{3}{c}{RACS-low} & \multicolumn{3}{c}{RACS-mid} \\
\hline
Name & Gaia J2016 & Epoch & Gaia sep (\arcsec) & Flux density (mJy) & Epoch & Gaia sep (\arcsec) & FIRST sep (\arcsec) & Epoch & Gaia sep (\arcsec) & FIRST sep (\arcsec) & Epoch & Gaia sep (\arcsec) & FIRST sep (\arcsec) \\ 
\hline
\hline
FIRST J093148.2$+$394833 & 9:31:48.3 $+$39:48:33.2 & J1994.6 & 0.15 & 1.23 & J2019.3 & 0.10 & 244.40 &  &  &  & J2021.0 & 0.11 & 6.76 \\
HD  77407 FIRST\,J090327.0$+$375029 & 9:03:27.0 $+$37:50:26.6 & J1994.6 & 0.96 & 1.67 & J2019.3 & 4.35 & 29.74 &  &  &  & J2021.0 & 4.64 & 8.15 \\
PM J11240$+$3808 FIRST\,J112404.3$+$380810 & 11:24:04.5 $+$38:08:10.7 & J1994.6 & 0.34 & 1.14 & J2019.3 & 3.02 & 371.00 &  &  &  & J2021.0 & 3.22 & 287.66 \\
V* AZ Psc FIRST\,J225852.9$-$001857 & 22:58:53.0 $-$0:18:57.2 & J1999.2 & 0.30 & 2.28 & J2018.0 & 1.08 & 132.27 & J2019.3 & 1.15 & 130.95 & J2021.0 & 1.25 & 132.10 \\
V* FP Cnc B FIRST\,J080855.4$+$324906 & 8:08:55.4 $+$32:49:01.4 & J1994.6 & 0.26 & 2.47 & J2019.3 & 5.30 & 390.49 &  &  &  & J2021.0 & 5.65 & 261.38 \\
V* MS Ser FIRST\,J155844.0$+$253411 & 15:58:43.9 $+$25:34:08.5 & J1995.9 & 0.45 & 2.28 & J2017.9 & 3.12 & 315.73 & J2020.3 & 3.46 & 84.38 & J2021.0 & 3.56 & 5.92 \\
\hline 
\end{longtable}
\end{small}
\end{landscape} 

\clearpage
\twocolumn

{We performed a simulation of steps \ref{step: cut Gaia pm var} to \ref{step: parallax cut var} to determine how likely it is that the 54 candidate variable radio stars are chance coincidence between the radio source and an optical source. We did this using FIRST, \gaia\, and RACS-mid. We took the positions of the FIRST sources and randomised their Right Ascension and Declination. We offset each source by taking the square root of a number drawn from a random uniform distribution between shift$^{2}$ and (shift$+$radius)$^{2}$ where shift$=$2\asec\, and radius$=$15\asec. This offset was in a direction chosen by selecting an angle between 0 and 360 from a random uniform distribution. We chose a minimum shift of 2\arcsec\,as our match radius is 1\arcsec\,and we did not want real matches in our random matches. We similarly randomised the positions of the RACS-mid sources using a shift$=$4\asec\, and radius$=$15\asec. We also needed to account for proper-motion. We selected a FIRST epoch by drawing from a random uniform distribution with a minimum and maximum matching the first and last FIRST epochs. We did the same to select a RACS-mid epoch. We then used the randomised FIRST and RACS-mid positions and epochs to perform steps \ref{step: cut Gaia pm var} to \ref{step: parallax cut var}. We performed this 50,000 times and recorded the number of resultant candidate variable stars per iteration. This resulted in a Poisson distribution with $\lambda=3.2$, where $\lambda$ is the expectation value and variance of the distribution. This means that it is likely that between 1 and 5 of the 54 matches are chance coincidence.}

 {Nine of the sources, marked by numbers in Table\,\ref{tab: candidate variable star literature}, have previously been identified as radio stars.} We searched the literature and archives to further investigate the candidate variable radio stellar sources. We searched each survey/catalogue within 1\arcsec\,of the FIRST position of each source. Many of the sources have been classified as optical counterparts to radio sources in the past, the results of the search are shown in Table\,\ref{tab: candidate variable star literature}. {In the table and in the descriptions here we have used ``stellar'' and ``star'' as in the original references. ``Stellar'' is used to mean a source with an unresolved or ``point-like'' point spread function (PSF), while a ``star'' is ``a self-luminous gaseous celestial body'' \citep{2020ApJS..249....3A}.}

{\citet{2001ApJS..135..227B} and \citet{2002ApJS..143....1M} matched FIRST sources to the Cambridge Automated Plate Measuring Machine (APM) scans of the POSS I plates. \citet{2001ApJS..135..227B} used a match radius of 1\farcs2 while \citet{2002ApJS..143....1M} used a match radius of 5\arcsec\ between FIRST and APM sources. In both of these surveys a source is classified as ``stellar'' if the optical source had an unresolved or ``point-like'' PSF in the relevant optical survey. \citet{2002ApJS..143....1M} performed extensive exploration of the chance coincidence and completeness of the matching and estimated that 98 per cent of APM sources within 1\arcsec\, of a FIRST source were physically associated.}


{45 of the sources were classed as matches to Sloan Digital Sky Survey \citep[SDSS;][]{2009ApJS..182..543A} ``stellar'' sources by \citet{2015ApJ...801...26H}. In the SDSS catalogue stellar sources are mostly quasars and AGN with a smaller fraction of radio stars.} They matched FIRST and SDSS using a 4\farcs8 match radius, they found that 19 per cent of FIRST-SDSS matches at a radius of 4\farcs8 are false/chance. 

Eight of the sources were classed as candidate radio stars by \citet{2009ApJ...701..535K}. They matched FIRST and SDSS using a match radius of $1\arcsec$. They also filtered the SDSS optical sources such that $r\leq20.5$ mag and the FIRST radio sources such that $S_{20}\geq1.25$ mJy (where $S_{20}$ is the FIRST flux density). Using these criteria they found that 98 per cent of their matches were physically associated. After matching, they filtered out quasars from the sample using SDSS spectra and literature investigations of the sources. They then visually inspected the radio sources and removed any resolved or complex morphology sources. However, they concluded that most if not all of their candidate stellar radio sources were actually chance alignments between the radio and optical. 

26 of the sources were classed as stars in the Million Optical--Radio/X-ray \citep[MORX;][]{2004A&A...427..387F,2016PASA...33...52F} Associations Catalogue. They used an algorithm described in \citet{2004A&A...427..387F} to calculate the likelihood of each match. The confidence of the match is included in the catalogue. 

42 of the sources match within $1\arcsec$ of at least one SDSS Date Release 16 \citep[SDSS DR16][]{2020ApJS..249....3A} source. SDSS also classes their sources as either a galaxy or a star. 28 of the sources have only stars within $1\arcsec$ of the FIRST position, while 14 have at least one source classed as a galaxy or unknown within $1\arcsec$ of the FIRST position. The matches to these catalogues for each of our 54 candidate variable radio stellar sources are shown in Table\,\ref{tab: candidate variable star literature}. The distances to these sources range from $\sim30$\,pc (HD  77407) to $\sim2300$\,pc (Gaia DR3 6910163884580689280).

\clearpage
\onecolumn

\clearpage
\onecolumn

\begin{landscape}
\centering
\begin{small}
\begin{longtable}{lrrrrrrrA{1.0cm}rA{1.5cm}rA{1.2cm}A{1.1cm}A{1.2cm}}
\caption[]{\label{tab: candidate variable star literature} {Literature classifications for the candidate variable radio star sources identified using FIRST and RACS-mid. We searched the FIRST position of each source with a 1\arcsec\,radius in each survey. ``Class'' indicates the classification of the source by that survey. The classes from \citet{2016PASA...33...52F} are ``S'' for star, ``R'' for radio, and ``X'' for X-ray. {\citet{2001ApJS..135..227B}, \citet{2002ApJS..143....1M} and \citet{2015ApJ...801...26H} use ``stellar'' to indicate that the PSF of the optical source is unresolved/point-like, which may include quasars/AGNs as well as stars.}``Radio conf'' indicates the confidence that the optical-radio match is physical. The separations are the separations between the radio source and the optical counterpart identified. The last three columns indicate whether there is a source classed as a galaxy, star or unknown by SDSS DR16 \citep[``Y'' for yes and ``N'' for no;][]{2020ApJS..249....3A}. Sources marked with a $\ddag\ddag$ were explored further as the literature search showed possible galaxy identification.}}\\
 & \multicolumn{2}{c}{\citet{2002ApJS..143....1M}} & \multicolumn{2}{c}{\citet{2001ApJS..135..227B}} & \multicolumn{2}{c}{\citet{2015ApJ...801...26H}} & \multicolumn{2}{c}{\citet{2009ApJ...701..535K}} & \multicolumn{3}{c}{\citet{2016PASA...33...52F}} & \multicolumn{3}{c}{\citet{2020ApJS..249....3A}} \\
Name & Sep (\arcsec) & class & Sep (\arcsec) & class & Sep (\arcsec) & class & Sep (\arcsec) & stellar class & class & radio conf & Sep (\arcsec) & galaxy within 1\arcsec & star within 1\arcsec & unknown within 1\arcsec \\
\hline
\hline 
2MASS J07500030+3458579 &0.9 & stellar &  &  & 0.92 & stellar & 0.85 & M1 & SR & 97.8 & 0 & N & Y & N \\
2MASS J08293480+0858099 & &  &  &  & 0.79 & stellar &  &  & SR & 97.3 & 0 & N & Y & N \\
2MASS J09244488+0019097 & &  &  &  & 0.94 & stellar & 0.9 & K7 &  &  &  & N & Y & N \\
2MASS J09420757+0334344$^{\ddag\ddag}$ & &  &  &  & 0.37 & galaxy &  &  & SR & 99.0 & 1 & Y & N & N \\
2MASS J09582770+2847572 &2.3 & stellar &  &  & 0.5 & stellar & 0.44 & M3 &  &  &  & N & Y & N \\
2MASS J10014486+2756455 &2.1 & stellar &  &  & 0.66 & stellar & 0.65 & M2 & SR & 89.1 & 1 & N & Y & N \\
2MASS J14333139+3417472$^{1}$ &4.1 & stellar &  &  & 1.5 & stellar &  &  &  &  &  &  &  &  \\
2MASS J15085996+2714307$^{\ddag\ddag}$ &0.43 & stellar &  &  & 0.62 & stellar &  &  & SR & 97.3 & 0 & Y & Y & N \\
2MASS J15215160+4246246 &1.3 & noise &  &  & 0.22 & stellar & 0.21 & M1 & SR & 98.4 & 1 & N & Y & N \\
2MASS J16234398+1302124 & &  &  &  & 0.91 & stellar &  &  & SR & 97.2 & 0 & N & Y & Y \\
2MASS J20485716-0053473 & &  &  &  & 0.57 & stellar &  &  &  &  &  & N & Y & N \\
BD+09  4984B$^{\ddag\ddag}$ & &  &  &  &  &  &  &  & RX & 99.0 & 0 & Y & N & N \\
BD-08  6022 & &  &  &  & 0.11 & stellar &  &  & SRX & 97.8 & 0 & N & Y & N \\
LP  521-15 & &  &  &  & 2.8 & stellar &  &  &  &  &  &  &  &  \\
Gaia DR3 1158730503710098688 & &  &  &  & 0.59 & stellar &  &  &  &  &  & N & Y & N \\
Gaia DR3 1360691101602998784 &0.72 & stellar &  &  & 0.58 & stellar &  &  & R & 99.2 & 0 & N & Y & N \\
Gaia DR3 1732181789609582208 & &  &  &  & 0.78 & stellar &  &  &  &  &  & N & Y & Y \\
Gaia DR3 2377310715263856512 & &  &  &  & 0.57 & stellar &  &  &  &  &  & Y & Y & N \\
Gaia DR3 2467183490048356864 & &  &  &  & 0.85 & stellar &  &  &  &  &  & N & Y & N \\
Gaia DR3 2490713524213973248 & &  &  &  & 0.51 & stellar &  &  &  &  &  & N & Y & N \\
Gaia DR3 2633166727449471360 & &  &  &  & 0.32 & stellar &  &  &  &  &  & N & Y & Y \\
Gaia DR3 2715431259726742784 & &  &  &  & 0.62 & stellar &  &  & R & 92.6 & 1 & N & Y & N \\
Gaia DR3 2776753256590751104 & &  &  &  & 0.78 & stellar &  &  &  &  &  & N & Y & N \\
Gaia DR3 3147019745776728448 & &  &  &  & 0.69 & stellar &  &  & SR & 72.0 & 1 & N & Y & N \\
Gaia DR3 3592258045911461376 & &  &  &  &  &  &  &  &  &  &  &  &  &  \\
Gaia DR3 6910163884580689280 & &  &  &  & 0.39 & stellar &  &  & SR & 71.1 & 1 & N & Y & N \\
SDSS J161007.07+394132.8 &0.5 & stellar &  &  & 0.47 & stellar &  &  & SR & 98.2 & 0 & N & Y & N \\
StKM 1-1155 & &  &  &  & 0.34 & stellar &  &  & SRX & 98.7 & 0 & N & Y & N \\
TYC 2503-1270-1$^{\ddag\ddag}$ &4.3 & blended &  &  & 0.36 & stellar &  &  & SR & 92.0 & 2 & Y & Y & N \\
UCAC4 431-063012 & &  &  &  &  &  &  &  &  &  &  &  &  &  \\
V* FF Aqr$^{\ddag,2}$ & &  &  &  & 0.54 & galaxy &  &  & SRX & 92.2 & 1 & Y & Y & N \\
V* IN Leo$^{\ddag\ddag}$ & &  &  &  & 0.47 & galaxy &  &  & SRX & 98.5 & 0 & Y & Y & N \\
V* V436 Ser$^{\ddag\ddag}$ & &  &  &  & 0.61 & stellar &  &  & SRX & 97.8 & 0 & Y & Y & N \\
Wolf  424 A$^{3}$ & &  &  &  & 6.0 & stellar &  &  & R & 98.7 & 0 &  &  &  \\
Wolf  424 B$^{3}$ & &  &  &  & 6.0 & stellar &  &  & R & 98.7 & 0 &  &  &  \\
$[$UBW2009$]$ 21 & &  &  &  & 0.98 & stellar &  &  &  &  &  & N & Y & N \\
Gaia DR3 1157257742244515584 & &  &  &  & 0.69 & stellar & 0.71 & G6 & SR & 80.2 & 1 & N & Y & N \\
Gaia DR3 1305574679646964352 &0.49 & non-stellar &  &  & 1.1 & stellar &  &  &  &  &  &  &  &  \\
2MASS J03065628+0044316$^{4}$ & &  &  &  & 0.48 & stellar &  &  & SR & 82.3 & 2 & N & Y & Y \\
2MASS J07415981+2331589$^{\ddag\ddag}$ &0.89 & stellar & 0.82 & stellar & 0.79 & stellar &  &  & SR & 86.1 & 1 & Y & Y & N \\
2MASS J16115068+4344126 &0.99 & stellar &  &  & 0.17 & stellar &  &  & SR & 79.2 & 1 & N & Y & N \\
2MASS J16235772+2350113 &1.3 & noise &  &  & 0.55 & stellar & 0.57 & M3 & SR & 89.7 & 1 & N & Y & N \\
FBQS J0748+3709 &0.59 & stellar & 0.52 & stellar & 0.68 & stellar &  &  &  &  &  & N & Y & N \\
FBQS J0754+3937 &0.88 & stellar & 0.33 & stellar & 0.62 & stellar & 0.58 & F8 & SR & 98.2 & 1 & N & Y & N \\
FBQS J1216+3020 &1.3 & non-stellar &  &  & 0.39 & stellar &  &  & R & 99.1 & 0 & N & Y & N \\
 & \multicolumn{2}{c}{\citet{2002ApJS..143....1M}} & \multicolumn{2}{c}{\citet{2001ApJS..135..227B}} & \multicolumn{2}{c}{\citet{2015ApJ...801...26H}} & \multicolumn{2}{c}{\citet{2009ApJ...701..535K}} & \multicolumn{3}{c}{\citet{2016PASA...33...52F}} & \multicolumn{3}{c}{\citet{2020ApJS..249....3A}} \\
Name & Sep (\arcsec) & class & Sep (\arcsec) & class & Sep (\arcsec) & class & Sep (\arcsec) & stellar class & class & radio conf & Sep (\arcsec) & galaxy within 1\arcsec & star within 1\arcsec & unknown within 1\arcsec \\
\hline
\hline
FBQS J1421+3319 &0.72 & stellar & 0.51 & stellar & 1.0 & stellar &  &  &  &  &  &  &  &  \\
FBQS J1704+2931 &0.43 & stellar & 0.43 & stellar & 0.56 & stellar &  &  & SR & 75.4 & 1 & N & Y & N \\
FBQS J1707+3802 &0.23 & stellar & 0.23 & stellar & 0.57 & stellar &  &  & SR & 82.1 & 1 & N & Y & N \\
FIRST J093148.2+394833 &0.25 & stellar &  &  & 0.11 & stellar &  &  & SR & 98.6 & 0 & N & Y & N \\
HD  77407$^{\ddag,5}$ &12 & non-stellar &  &  & 2.5 & galaxy &  &  &  &  &  &  &  &  \\
PM J11240+3808 &4.6 & stellar &  &  & 1.2 & stellar &  &  &  &  &  &  &  &  \\
V* AZ Psc$^{\ddag,5}$ & &  &  &  & 0.14 & galaxy &  &  & SRX & 57.6 & 1 & Y & N & N \\
V* FP Cnc B$^{5}$ &18 & stellar &  &  & 1.7 & stellar &  &  &  &  &  &  &  &  \\
V* MS Ser$^{\ddag,5}$ &9.1 & non-stellar &  &  & 2.3 & galaxy &  &  &  &  &  &  &  &  \\
    \hline
\end{longtable}
\begin{tablenotes}
      \small
      \item (1) identified as a radio star using circular polarisation by \citet{2021NatAs.tmp..196C}
      \item (2) identified as a radio star by \citet{1988AJ.....95..204M}
      \item (3) identified as a radio flaring star by \citet{1974ApJ...190L.129S}
      \item (4) identified as a radio star by \citet{2001AJ....122..518H}
      \item (5) identified as a radio star by \citet{1999AJ....117.1568H}
    \end{tablenotes}
\end{small}
\end{landscape} 

\clearpage
\twocolumn

{We used the literature search to investigate some sources in more detail. In particular, we checked those sources that were classed as galaxies by \citet{2015ApJ...801...26H} and those sources where there is a galaxy within 1\arcsec\, as determined by \citet{2020ApJS..249....3A}. Some of these sources (BD+09  4984B, TYC 2503-1270-1, V* FF Aqr, V* IN Leo, V* V436 Ser, HD  77407, V* AZ Psc, V* MS Ser) are well-known stars that were mis-classified as galaxies as they are extremely bright optical sources. Two sources (Gaia DR3 2377310715263856512 and 2MASS J07415981$+$2331589) are unlikely to be radio stars. Gaia DR3 \\2377310715263856512 and 2MASS J07415981$+$2331589 are less than 2\arcsec\, from faint optical sources that do not have parallax or proper-motion measurements. These nearby optical sources may be galaxies and as such we find it more likely that the radio emission is associated with these galaxies. We conclude that we have found nine previously known variable radio stars and $43$ candidate variable radio stars; where 5 candidates are likely to be chance coincidence.
}

\section{Discussion}
\label{sec: discussion}

We have found eight radio stars using their proper-motion, two of which (PM J15587$+$2351E and GS Leo) have not previously been identified as radio stars. {We have also found $43$ candidate variable radio stellar sources and nine known radio stars.}

The set of proper-motion radio stars is likely volume limited. This is because the further the star is from Earth, the further the star needs to travel to achieve a high proper-motion (all of the following proper-motion calculations are in two dimensions). In the set of RACS-mid detected stars, sig Gem travels the furthest: 0.001\,pc.
The smallest $D_\mathrm{{G_{FIRST,RM}}}$ value for RACS-mid detected stars is 39 Cet: 2.71\arcsec.
The distance at which 0.001\,pc$\approx2.71\arcsec$ is $\sim 90$ pc. This provides an approximate maximum distance that we have probed by searching for FIRST--RACS-mid proper-motion radio stars. In the set of VLASS detected stars, BI Cet travels the furthest: 0.002\,pc, while the smallest $D_\mathrm{{G_{FIRST,RM}}}$ value for VLASS detected stars is FK Com: $1.23\arcsec$. The distance at which 0.002\,pc$\approx1.23\arcsec$ is $\sim250$ pc. This provides an approximate maximum distance that we have probed by searching for FIRST--VLASS proper-motion radio stars. This means that we can increase the volume we are probing by increasing the time baseline between radio surveys, or decreasing the position uncertainties of the radio surveys, or both. Sensitivity also plays an important role in the volume we can probe to. \citet{2021MNRAS.502.5438P} detected stellar radio sources out to a distance of $150$ pc using RACS-low, the RS CVn-like system $\mathrm{MKT\,J170456.2-482100}$ detected by MeerKAT in ten minute images is $550$ pc away \citep[][]{2020MNRAS.491..560D}, and in the set of 8 stellar sources presented here, FK Com is detected by FIRST, VLASS and RACS-mid at $\sim220$ pc. New instruments such as the SKA will be even more sensitive, and therefore will be able to detect stellar sources beyond $250$ pc.

To find stars using the proper-motion method we require stars that either have persistent radio emission or are serendipitously flaring in both radio epochs, as well as reasonably high proper-motion.
This means that we do not know the expected number of stars that could be detected using the various current and future radio surveys.
However, we can use the eight radio stars we have detected with this method to determine the time baseline required for various radio surveys to detect those stars. Of the eight stars we found, FK Com has the lowest proper-motion (56.615 mas yr$^{-1}$) and sig CrB A has the highest proper-motion (282.061 mas yr$^{-1}$).
in Table\,\ref{tab: survey time baselines} we show the minimum time baseline required to find these stars, where we assume the star will be found when $D_\mathrm{{G_{A,B}}}>a\arcsec+b\arcsec$. MeerKAT has a similar astrometric precision to FIRST, so we have only included FIRST in the table. The astrometric precision values for SKA-low and SKA-mid are the resolutions listed in \citet{2019arXiv191212699B}.
We can see from Table\,\ref{tab: survey time baselines} that we would only require a {time} baseline of $\sim0.6$ years with the SKA-mid at 6.7 {GHz} to use this method to find a source with the same proper-motion as Sig CrB A. If we assume that SKA will be operational in 2030 then, of the eight proper-motion stars, BI Cet will have travelled the furthest since its J1998.8 FIRST detection: $\sim0.0025$ pc (8.3\arcsec\,at a distance of 62.12 pc). If we assume that the minimum angular separation between the FIRST position of a source and the SKA 1.4 GHz position of a source is 1.4\arcsec\,then the maximum distance a source that travelled $\sim0.0025$ pc could be detected at is $\sim370$ pc. This probes $\sim1.5$ times the distance probed by FIRST--VLASS and $\sim4$ times the distance probed by FIRST--RACS-mid. If we assume that the minimum angular separation between the FIRST position of a source and the SKA 6.7 GHz position of a source is 1.08\arcsec\,then the maximum distance a source that travelled $\sim0.0025$ pc could be detected at is $\sim480$ pc. This probes $\sim2$ times the distance probed by FIRST--VLASS and $\sim5$ times the distance probed by FIRST--RACS-mid. Two sky surveys using SKA 6.7 GHz would probe to a similar distance if the two surveys were performed $\sim4.5$ years apart, assuming that the sensitivity to stellar sources (taking into account radio spectral indices and SKA sensitivity at 6.7 GHz) is comparable to 1.4 GHz. \citet{2019arXiv191212699B} suggests that the resolution of SKA-mid at 12.5 GHz will be 0.04\arcsec. If we again assume that the sensitivity to stellar sources is comparable to 1.4 GHz, two surveys performed by SKA-mid at 12.5 GHz would probe to $\sim480$ pc if the two surveys were performed $\sim2.5$ years apart.
Even proper-motion matching between the lowest precision survey (RACS) and the highest precision survey (SKA-mid) requires less than 10 years to find sources with similar proper-motions to sig CrB A. This demonstrates how the SKA will expand our searches for stellar radio sources.
MeerKAT L-band observations have a similar astrometric accuracy to FIRST, which is important because MeerKAT is in the Southern hemisphere, compared to FIRST in the Northern hemisphere.
This means that both RACS and MeerKAT will be key for providing the early-time observations to initially compare to SKA-mid.

{The LOFAR Two-metre Sky Survey \citep[LOTSS;][]{2017A&A...598A.104S,2022A&A...659A...1S} catalogue} does not currently provide the epoch of detection for each source because they revisit each field multiple times to achieve higher sensitivity and uv-coverage. However, for projects where proper-motion is important, it is key that epoch of detection/observation is provided.
This can be done using an average epoch, similar to what is provided by FIRST, or by providing two catalogues: a catalogue of sources extracted from the deep/stacked images with no epoch provided, and a catalogue of sources extracted from single epoch observations with the epoch included.
Many radio sources are persistent and extra-galactic, which means that having accurate epoch information is not required for those sources.

It is now possible to perform large-scale searches for stellar radio sources \citep[e.g.][]{2021MNRAS.502.5438P,2021NatAs...5.1233C} where proper-motion is important, plus searches for variable and transient radio sources (e.g. Variables and Slow Transients with ASKAP\footnote{\href{https://www.vast-survey.org/}{https://www.vast-survey.org/}} \citep[VAST;][]{2013PASA...30....6M}, ThunderKAT\footnote{\href{http://www.thunderkat.uct.ac.za/}{http://www.thunderkat.uct.ac.za/}} \citep{2017arXiv171104132F}. 
We also need epoch information to account for sources with proper-motion when we are searching for variable and transient radio sources on long time scales.
This is because we could identify a source as transient when it has just moved across the sky. This will be particularly important for high-resolution instruments such as the SKA and even current surveys with LOFAR, as the time baseline required for a star to move out of the position uncertainty region of its first detected position can be as small as $\lesssim1$ year. For example, two of the candidate radio variable stellar sources are the components of the the binary system Wolf 424 (see Table\,\ref{tab: variable star cands}). These stars have proper-motions magnitudes of 1.7 arcsec $\mathrm{yr^{-1}}$ and 1.8 arcsec $\mathrm{yr^{-1}}$. This means that they would be detected as two separate transients by VLASS, LOTSS and the SKA when comparing observations observed less than six months apart. Even with RACS, the survey with the highest position uncertainties, these stars would be identified as two radio transients in less than two years.
Conversely, we may misidentify a transient source as extra-galactic if we do not account for the positions of stellar sources at each epoch.
As such, we need to consider epoch information and how to appropriately include it in all current and future radio sky surveys and databases.

\begin{table*}
\caption{Time baselines required when performing proper-motion searches between current and future radio sky surveys. t$_{\mathrm{min}}$ is the time it would take to use the proper-motion method to find sig CrB A, the source presented in Section\,\ref{sec: demo results} with the highest proper-motion. t$_{\mathrm{max}}$ is the time it would take to use the proper-motion method to find FK Com, the source presented in Section\,\ref{sec: demo results} with the lowest proper-motion. MeerKAT has a similar astrometric precision to FIRST, $\sim1$\arcsec, so we have only included FIRST in the table. SKA 0.77 is SKA-low at 770 MHz, SKA 1.4 is SKA-mid at 1.4 GHz, and SKA 6.7 in SKA-mid at 6.7 GHz \citep{2019arXiv191212699B}.}
    \centering
    \begin{tabular}{lrrrrrrrrrrrrrr}
    \hline
     & \multicolumn{2}{c}{FIRST (1\arcsec)} &  \multicolumn{2}{c}{RACS (2\arcsec)} &  \multicolumn{2}{c}{VLASS (0.5\arcsec)} &  \multicolumn{2}{c}{LOTSS (0.2\arcsec)} &  \multicolumn{2}{c}{SKA 0.77 (0.7\arcsec)} &  \multicolumn{2}{c}{SKA 1.4 (0.4\arcsec)}  &  \multicolumn{2}{c}{SKA 6.7 (0.08\arcsec)} \\
     & t$_{\mathrm{min}}$  & t$_{\mathrm{max}}$  & t$_{\mathrm{min}}$  & t$_{\mathrm{max}}$  & t$_{\mathrm{min}}$ & t$_{\mathrm{max}}$  & t$_{\mathrm{min}}$ & t$_{\mathrm{max}}$  & t$_{\mathrm{min}}$ & t$_{\mathrm{max}}$ & t$_{\mathrm{min}}$  & t$_{\mathrm{max}}$& t$_{\mathrm{min}}$  & t$_{\mathrm{max}}$  \\
    \hline
    FIRST (1.0\arcsec) & \textbf{7.1yr} & \textbf{35.3yr} & 10.6yr & 53.0yr & 5.3yr & 26.5yr & 4.3yr & 21.2yr & 6.0yr & 30.0yr & 5.0yr & 24.7yr & 3.8yr & 19.1yr \\
    RACS (2.0\arcsec) & 10.6yr & 53.0yr & \textbf{14.2yr} & \textbf{70.7yr} & 8.9yr & 44.2yr & 7.8yr & 38.9yr & 9.6yr & 47.7yr & 8.5yr & 42.4yr & 7.4yr & 36.7yr \\
    VLASS (0.5\arcsec) & 5.3yr & 26.5yr & 8.9yr & 44.2yr & \textbf{3.5yr} & \textbf{17.7yr} & 2.5yr & 12.4yr & 4.3yr & 21.2yr & 3.2yr & 15.9yr & 2.1yr & 10.2yr \\
    LOTSS (0.2\arcsec) & 4.3yr & 21.2yr & 7.8yr & 38.9yr & 2.5yr & 12.4yr & \textbf{1.4yr} & \textbf{7.1yr} & 3.2yr & 15.9yr & 2.1yr & 10.6yr & 1.0yr & 4.9yr \\
    SKA 0.77 (0.7\arcsec) & 6.0yr & 30.0yr & 9.6yr & 47.7yr & 4.3yr & 21.2yr & 3.2yr & 15.9yr & \textbf{5.0yr} & \textbf{24.7yr} & 3.9yr & 19.4yr & 2.8yr & 13.8yr \\
    SKA 1.4 (0.4\arcsec) & 5.0yr & 24.7yr & 8.5yr & 42.4yr & 3.2yr & 15.9yr & 2.1yr & 10.6yr & 3.9yr & 19.4yr & \textbf{2.8yr} & \textbf{14.1yr} & 1.7yr & 8.5yr \\
    SKA 6.7 (0.08\arcsec) & 3.8yr & 19.1yr & 7.4yr & 36.7yr & 2.1yr & 10.2yr & 1.0yr & 4.9yr & 2.8yr & 13.8yr & 1.7yr & 8.5yr & \textbf{0.6yr} & \textbf{2.8yr} \\
    \hline
    \end{tabular}
    \label{tab: survey time baselines}
 \end{table*}

{We found eight radio stellar sources using proper-motion searching, but we found $43$ candidate variable stellar sources and nine known radio stars (for a total of $52$ sources, where five sources are likely chance coincidence) when we expanded our search to variable radio sources at $\sim1400$\,MHz.
The large number of variable sources compared to the small number of sources detected in two epochs implies that most stellar radio sources are detected because they flare, not because they are persistently bright in the radio. 
The area of the sky where RACS-mid and FIRST overlap is $-11.5 < \mathrm{declination} < 49.5$ which gives a solid angle of $\sim20\,000$ deg$^{2}$. 
The six radio stellar sources found using FIRST--RACS-mid proper-motion searching range from $\sim22$ pc to $\sim75$ pc away, resulting in a surface density of $3\times10^{-4}$ deg$^{-2}$ or a density of $2\times10^{-9}$ pc$^{-3}$. The $52$ variable stellar sources (9 known and 43 candidate) range from $\sim30$ pc to $\sim2300$ pc away, resulting in a surface density of $2.6\times10^{-3}$ deg$^{-2}$ or a density of $2.6\times10^{-10}$ pc$^{-3}$.
So while we probe deeper searching for variable stars, there is a higher number of stars per volume found using the proper-motion method.}

\begin{figure*}
    \caption{\label{fig: BI Cet variability} ASKAP images of BI Cet. These images are from the available observations of BI Cet in CASDA, including RACS-low and VAST observations.
    The epochs and frequencies of the observations are shown on the images. Note that these observations have various integration times.
    The markers are the same as those presented in Figure\,\ref{fig: star radio images}}
        \begin{subfigure}{\textwidth}
            \includegraphics[width=0.95\textwidth]{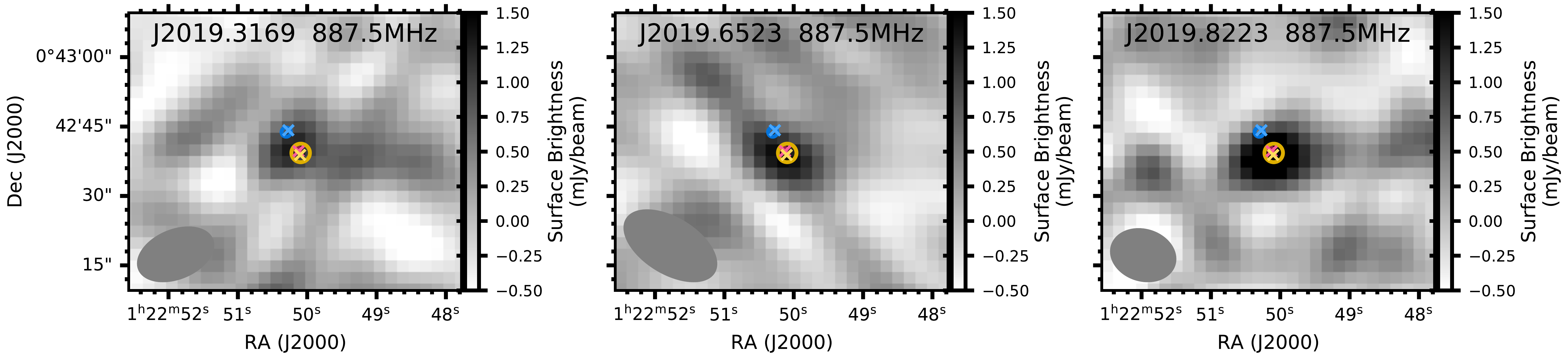}
        \end{subfigure}
        \begin{subfigure}{\textwidth}
            \includegraphics[width=0.95\textwidth]{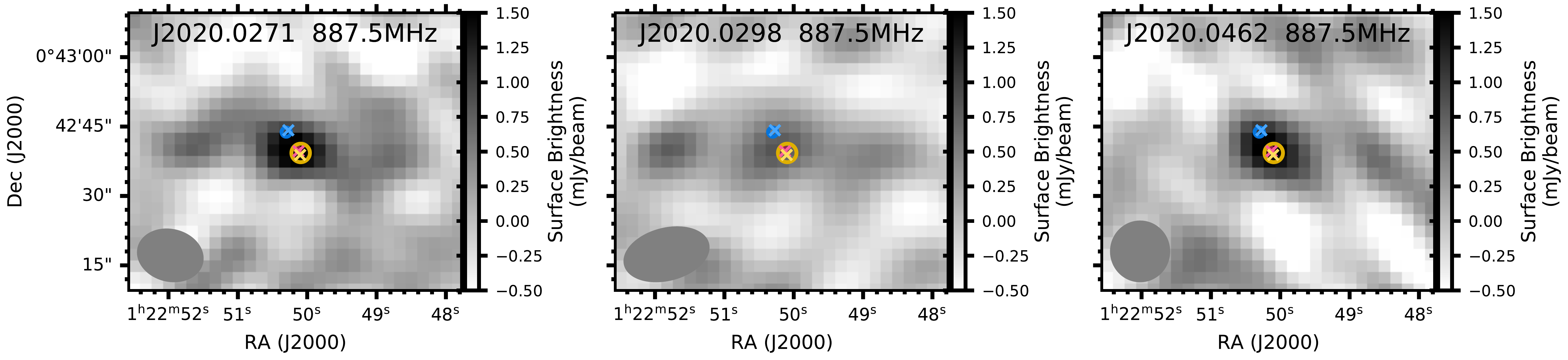}
        \end{subfigure}
        \begin{subfigure}{\textwidth}
            \includegraphics[width=0.95\textwidth]{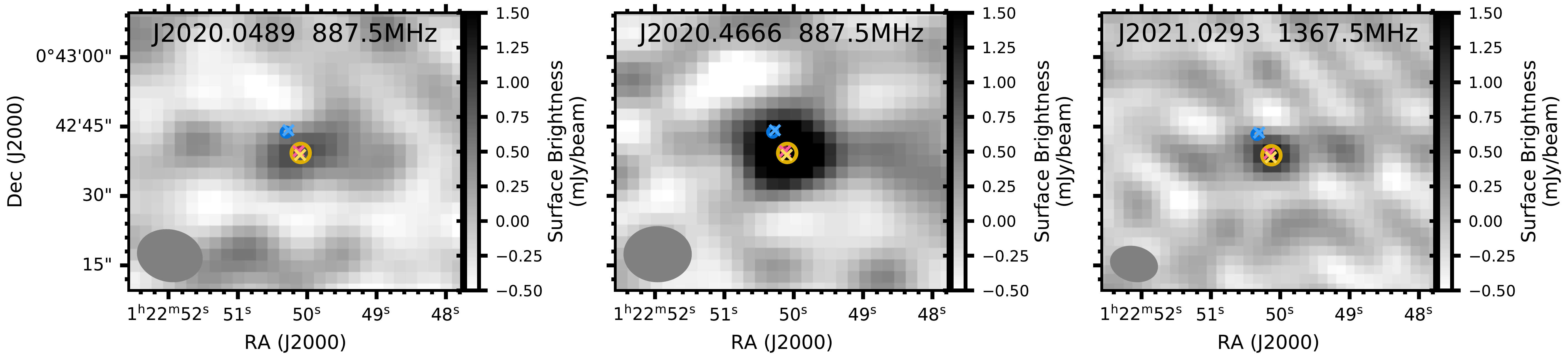}
        \end{subfigure}
        \begin{subfigure}{\textwidth}
            \includegraphics[width=0.95\textwidth]{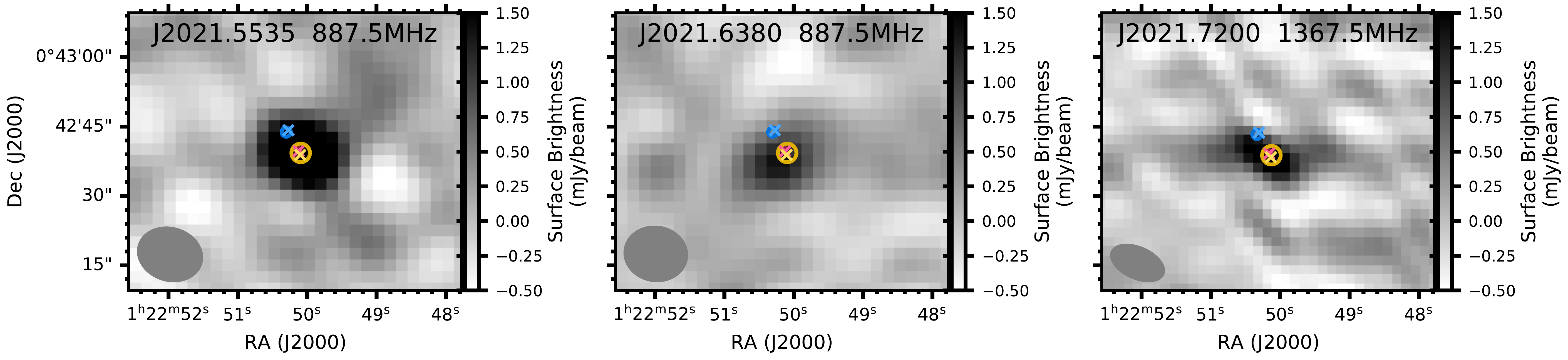}
        \end{subfigure}
\end{figure*}

This work demonstrates the star-finding power of performing high resolution all sky radio surveys with radio interferometers. Even with a single all-sky survey per instrument we can find both proper-motion stellar sources and candidate variable radio stellar sources. However, we found many more candidate variable radio stellar sources than persistent proper-motion sources. BI Cet is an excellent example of why performing multi-epoch sky surveys is important for stellar searches. BI Cet has been observed multiple times by ASKAP as part of various surveys. We can see in Figure\,\ref{fig: BI Cet variability} that its brightness varies significantly in the different ASKAP epochs. If only one ASKAP sky survey was performed on J2020.05, BI Cet would not have been detected. This means some stars may be missed because they are faint/quiescent in one or both radio surveys. Specialist survey instruments such as ASKAP can survey the whole sky in a matter of weeks. Repeat sky surveys on various time baselines could assist in finding more stellar sources that are only sporadically radio bright.

Stokes V searches 
are biased towards coherent radio emission mechanisms; while
variability searches are biased towards flaring stellar sources. Proper-motion searching only requires that the source is detected in two epochs. This means that it is biased towards sources that happened to flare in the two epochs or are persistently bright in the radio, but does not require further data on top of the radio continuum images and catalogues that are standard outputs of radio sky surveys. It also does not require a specific physical connection between the radio emission and the star to confirm that the star is the source of the radio emission. This is useful in reducing the biases in our searches for stellar radio sources. Combining the results of these search methods (Stokes V searches, variability searches, and proper-motion searches) is essential for finding as complete a sample of stellar radio sources as possible.

\section{Conclusions}
\label{sec: conclusions}

We have presented a method for identifying stellar radio sources using their proper-motion. We demonstrated this method using FIRST, VLASS, and RACS, and astrometric information from \gaia\, DR3; finding eight stellar radio sources, two of which had not previously been identified as radio stars. {We also found $43$ variable radio stars and nine known radio stars by searching for sources that were detected in FIRST that are not detected in RACS-mid.} Both of these methods will be important tools for identifying stellar radio sources as we perform sky surveys with existing instruments and plan for sky surveys with the SKA. In particular, we should endeavour to include epoch information in radio sky survey catalogues and consider survey strategies where each pointing of the sky is observed more than once.

\begin{acknowledgement}
This scientific work uses data obtained from Inyarrimanha Ilgari Bundara / the Murchison Radio-astronomy Observatory. We acknowledge the Wajarri Yamaji People as the Traditional Owners and native title holders of the Observatory site. CSIRO’s ASKAP radio telescope is part of the \href{https://ror.org/05qajvd42}{Australia Telescope National Facility}. Operation of ASKAP is funded by the Australian Government with support from the National Collaborative Research Infrastructure Strategy. ASKAP uses the resources of the Pawsey Supercomputing Research Centre. Establishment of ASKAP, Inyarrimanha Ilgari Bundara, the CSIRO Murchison Radio-astronomy Observatory and the Pawsey Supercomputing Research Centre are initiatives of the Australian Government, with support from the Government of Western Australia and the Science and Industry Endowment Fund.
This paper includes archived data obtained through the CSIRO ASKAP Science Data Archive, CASDA\footnote{\href{http://data.csiro.au/}{http://data.csiro.au/}}.
This work has made use of data from the European Space Agency (ESA) mission
{\it Gaia} (\href{https://www.cosmos.esa.int/gaia}{https://www.cosmos.esa.int/gaia}), processed by the {\it Gaia}
Data Processing and Analysis Consortium (DPAC,
\href{https://www.cosmos.esa.int/web/gaia/dpac/consortium}{https://www.cosmos.esa.int/web/gaia/dpac/consortium}). Funding for the DPAC
has been provided by national institutions, in particular the institutions
participating in the {\it Gaia} Multilateral Agreement.
This research made use of Astropy,\footnote{\href{http://www.astropy.org}{http://www.astropy.org}} a community-developed core Python package for Astronomy \citep{2013A&A...558A..33A,2018AJ....156..123A}.
This research made use of APLpy, an open-source plotting package for Python \citep{2012ascl.soft08017R}.
This research has made use of the VizieR catalogue access tool, CDS, Strasbourg, France\footnote{\href{10.26093/cds/vizier}{10.26093/cds/vizier}}. The original description 
of the VizieR service was published in \citet{2000A&AS..143...23O}.
This research has made use of the SIMBAD database,
operated at CDS, Strasbourg, France \citep{2000A&AS..143....9W}.
This research has made use of the CIRADA cutout service\footnote{\href{http://cutouts.cirada.ca/}{http://cutouts.cirada.ca/}}, operated by the Canadian Initiative for Radio Astronomy Data Analysis (CIRADA). CIRADA is funded by a grant from the Canada Foundation for Innovation 2017 Innovation Fund (Project 35999), as well as by the Provinces of Ontario, British Columbia, Alberta, Manitoba and Quebec, in collaboration with the National Research Council of Canada, the US National Radio Astronomy Observatory and Australia’s Commonwealth Scientific and Industrial Research Organisation.
This research has made use of NASA’s Astrophysics Data System Bibliographic Services\footnote{\href{https://ui.adsabs.harvard.edu/}{https://ui.adsabs.harvard.edu/}}.
LND would like to acknowledge the traditional owners of the land where most of her work was performed: the Wurundjeri People of the Woi worrung Nation, the Whadjuk people of the Noongar Nation and the Gadigal people of the Eora Nation.
We would like to thank the referee for their helpful comments.
\end{acknowledgement}


\bibliography{stars}



\end{document}